\documentclass[a4paper,fleqn,compress]{cas-dc}

\usepackage{geometry}
\usepackage{amsthm}
\theoremstyle{definition}
\newtheorem{definition}{Definition}

\usepackage{graphicx} 
\input{insbox}

\usepackage{lineno}   
\usepackage{lipsum}   
\usepackage{cancel, fancyhdr}
\usepackage{siunitx}  

\usepackage{amsmath,amssymb,amsfonts} 
\makeatletter
\def\ps@pprintTitle{%
  \let\@oddhead\@empty
  \let\@evenhead\@empty
  \let\@oddfoot\@empty
  \let\@evenfoot\@empty
  \renewcommand{\headrulewidth}{0pt}
  \renewcommand{\footrulewidth}{0pt}
}

\let\oldps@plain\ps@plain
\def\ps@plain{\oldps@plain} 
\makeatother

\usepackage{orcidlink}    
\usepackage{fontawesome5} 
\newcommand{\orcidicon}[1]{\href{https://orcid.org/#1}{\textsuperscript{\faOrcid}}}

\usepackage[T1]{fontenc}

\makeatletter
\renewcommand\section{%
  \@startsection{section}{1}{\z@}%
    {12\p@ \@plus 6\p@ \@minus 3\p@}
    {3\p@  \@plus 6\p@ \@minus 3\p@}
    {%
      \normalfont
      \bfseries                          
      \fontsize{10.8pt}{13pt}\selectfont 
    }%
}
\makeatother

\makeatletter
\renewcommand\subsection{%
  \@startsection{subsection}{2}{\z@}%
    {12\p@ \@plus 6\p@ \@minus 3\p@}
    {3\p@  \@plus 6\p@ \@minus 3\p@}
    {%
      \normalfont
      \fontsize{10.8pt}{13pt}\selectfont 
      \itshape                       
    }%
}

\renewcommand\subsubsection{%
  \@startsection{subsubsection}{3}{\z@}%
    {12\p@ \@plus 6\p@ \@minus 3\p@}
    {\p@}
    {\normalfont\normalsize\itshape}
}
\makeatother

\usepackage{xurl}
\usepackage{hyperref}
\hypersetup{
    colorlinks = true,
    breaklinks = true
}

\AtBeginDocument{%
    \hypersetup{
        linkcolor = cyan,
        urlcolor = cyan,
        citecolor = cyan,
        filecolor = cyan,
        allcolors = cyan
    }
}
\usepackage[numbers,sort&compress]{natbib}
\setcitestyle{square,comma,sort&compress}

\newlength\mylen
\setcitestyle{citesep={,\kern-\mylen}}

\usepackage{url}

\usepackage[nameinlink]{cleveref} 

\crefname{section}{Section}{Sections}
\Crefname{section}{Section}{Sections}

\crefname{figure}{Fig.}{Figs.}
\Crefname{figure}{Fig.}{Figs.}

\crefname{table}{Table}{Tables}
\Crefname{table}{Table}{Tables}

\newcounter{mainappendix}
\crefname{mainappendix}{Appendix}{Appendices}
\Crefname{mainappendix}{Appendix}{Appendices}
\crefformat{mainappendix}{#2Appendix#3}
\Crefformat{mainappendix}{#2Appendix#3}

\crefname{equation}{Eq.}{Eqs.}
\Crefname{equation}{Eq.}{Eqs.}

\usepackage{enumitem}
\SetEnumitemKey{prop}{
    label=\textbf{Property \arabic*}.,
    ref=Property \arabic*,
    leftmargin=*
}
\crefname{enumi}{Property}{Properties}
\Crefname{enumi}{Property}{Properties}

\SetEnumitemKey{step}{%
  label=\textbf{Step \arabic*}.,
  ref=Step \arabic*,
  leftmargin=*
}
\crefname{step}{Step}{Steps}
\Crefname{step}{Step}{Steps}

\usepackage[ruled]{algorithm2e} 

\SetInd{1em}{0.75em} 

\makeatletter
\renewcommand{\@algocf@post@ruled}{%
  \hrule height \algoheightrule\relax 
}

\SetAlCapHSkip{0pt} 
\SetAlgoCaptionLayout{textleft} 
\SetAlgoCaptionSeparator{:} 
\makeatother

\SetAlgoInsideSkip{medskip} 
\RestyleAlgo{ruled}

\crefname{algocf}{Algorithm}{Algorithms}
\Crefname{algocf}{Algorithm}{Algorithms}

\SetKwComment{Comment}{/* }{ */}

\usepackage{booktabs,threeparttable,tabularx,tablefootnote}
\usepackage{caption,subcaption}

\usepackage{float}          
\usepackage{multirow}       

\usepackage{picture, graphicx, graphics, eso-pic, tikz}

\usepackage{hyphenat}

\makeatletter
\renewcommand*\env@matrix[1][\arraystretch]{%
  \edef\arraystretch{#1}%
  \hskip -\arraycolsep
  \let\@ifnextchar\new@ifnextchar
  \array{*\c@MaxMatrixCols c}}
\makeatother

\usepackage{wrapfig} 

\begin{document}
\let\WriteBookmarks\relax
\def\floatpagepagefraction{1}
\def\textpagefraction{.001}
\let\printorcid\relax 

\title[mode = title]{DISPCA : A hybrid iterative-sequential approach for the identification of errors-in-variables model of linear DAE systems}

\author[1]{Deepanjhan Das}[style=chinese]

\author[2]{Vishwesh Ramanathan}[style=chinese]

\author[1]{Shankar Narasimhan\corref{cor1}}[style=chinese]%
\ead{naras@iitm.ac.in}

\cortext[cor1]{Corresponding author}





\affiliation[1]{
    organization={Department of Chemical Engineering, Indian Institute of Technology Madras},
    city={Chennai},
    postcode={600036},
    country={India}
}
\affiliation[2]{
    organization={University of Toronto},
    city={Toronto},
    postcode={M5S 1A1},
    country={Canada}
}


\begin{abstract}
The dynamic behavior of numerous engineering processes is effectively characterized through differential-algebraic equations (DAEs), commonly referred to as descriptor systems. While substantial progress has been achieved in identifying dynamic models governed by ordinary differential equations (ODEs), limited research has addressed the identification of descriptor systems from measured data. This work presents a systematic methodology for identifying the DAE model of a linear descriptor system in discrete difference equation form under errors-in-variables (EIV) setting, where both input and output measurements are corrupted by random noise. The proposed methodology generalizes the identification framework to handle scenarios where the system contains multiple algebraic and different ordered differential relations. The key innovation involves a partial stacking procedure of lagged data matrix with a sequentially increasing lag window that identifies all the differential relations individually. This is preceded by an iterative estimation of the measurement error covariance matrix that is diagonal and heteroskedastic, under large sample conditions. The algorithm simultaneously estimates the number of differential and algebraic relations, observability indices and delay parameters of the differential equations, and all the model coefficients directly from measured data without requiring prior specification from the user. The framework addresses the increased complexity arising from multiple dynamic coupled interactions while maintaining computational tractability through systematic decomposition of the identification problem. Effectiveness of the proposed methodology is demonstrated through several simulation studies. 
\end{abstract}


\begin{keywords}
Differential-algebraic equations \sep 
Linear descriptor systems \sep
Errors-in-variables models \sep
Partial stacking procedure \sep 
Principal component analysis \sep
System identification
\end{keywords}

\maketitle
\thispagestyle{plain}
\pagestyle{plain}
\vspace*{-\headsep} 


\section{Introduction} \label{sec:intro}
Differential-algebraic equations (DAEs), widely known as descriptor system, represent a powerful and fundamental class of mathematical models that have become increasingly prevalent in describing the dynamic behavior of complex engineering systems. Unlike the system of ordinary differential equations (ODEs), DAEs simultaneously incorporate both differential equations governing the dynamic evolution of system states and algebraic equations representing instantaneous constraints such as conservation laws, mechanical constraints, and equilibrium conditions \cite{Kunkel:2024,Montanari:2024}. This unique mathematical structure makes DAEs particularly well-suited for modeling constrained dynamical systems across diverse engineering domains, including chemical processes \cite{Montanari:2024,Rao:2003}, electrical circuits \cite{Kunkel:2024,Lena:2008}, biological systems \cite{Zhang:2012}, power networks \cite{Susuki:2008,Chakrabortty:2012}, and mechanical systems \cite{Duan:2010,Montanari:2024}. For instance, chemical reactors operating under material balance constraints, distillation columns with equilibrium relationships, and heat exchangers with energy conservation laws all naturally give rise to DAE formulations. 

Due to such high relevance, in last few decades substantial efforts and significant contributions from both classical and modern approaches have been made to extend the results of classical control to descriptor system \cite{Dai:1989,Duan:2010,Xu:2006}. Early foundational work established the theoretical framework for understanding DAE systems and their unique mathematical properties, such as the singular nature of the system matrices and the index structure \cite{Campbell:1995,Brenan:1995}, which differ significantly from conventional ODE systems \cite{Petzold:1982,Dai:1989,Brenan:1995,Campbell:1995}. Luenberger's \textit{shuffle algorithm} \cite{Luenberger:1978} represented one of the first systematic approaches to handle linear time-invariant (LTI) DAE systems. By providing a structured way to perform index reduction and transform higher-index systems into recursive forms, this method established the necessary groundwork for making such systems tractable for further analysis and identification purposes.

Transformation-based approaches emerged as primary methodological direction for DAE systems, converting DAE systems into state-space forms for conventional identification. Gerdin et al.~\cite{Gerdin:2007} established a theoretical foundation by transforming descriptor systems using oblique projections and canonical forms, enabling adaptation of state-space identification while preserving physical interpretability of the original DAE structure. The transformation approach also includes frequency domain methods. This has been proved particularly valuable for DAE systems where time-domain approaches encountered numerical difficulties, such as the direct inversion of ill-conditioned mass matrices during the identification process \cite{Pintelon:2001,Zhou:2024}.

A distinct and practically significant class of methods that parallels the transformation-based approaches is subspace-based identification, whose foundational application to standard linear state-space models has been extended to encompass descriptor systems. The pioneering contribution by Moonen et al.~\cite{Moonen:1992} established a subspace identification algorithm specifically tailored for descriptor models, projecting future outputs onto appropriate input-output subspaces to infer state sequences, from which both differential and algebraic components are subsequently extracted through least-squares estimation. Complementary advancements by Verhaegen and DeWilde~\cite{Verhaegen:1992}, and Van Overschee and De Moor~\cite{Van:1994} further consolidated subspace-based identification for general linear systems, offering computationally efficient algorithms that operate directly on input-output data without requiring explicit iterative optimization. While these two methodological directions; transformation and subspace-based; represent substantial progress, they share a critical underlying assumption that is the input measurements are assumed to be free of noise.

Similarly, the classical discrete-time identification approaches encompassing the prediction error method (PEM) \cite{Soderstrom:1989,Ljung:2002,Tangirala:2015} and classical output error maximum-likelihood (ML) estimators~\cite{Astrom:1980} yield consistent estimates under the assumption that all stochastic disturbances enter only through process or output channels. For linear descriptor systems specifically, the extension of ML estimators by Gerdin et al.~\cite{Gerdin:2007} offers statistical consistency and asymptotic optimality; however, such methods require careful handling of algebraic constraints, necessitating specialized algorithms to maintain numerical stability while ensuring constraint satisfaction. Recent developments have further extended these ideas to nonlinear DAE systems with process disturbances~\cite{Abdalmoaty:2021,Robert:2022}. Despite their theoretical merits, all these methods are restricted to scenarios where the input is treated as a deterministic or otherwise noise-free signal. 

\begin{figure}[!htb]
    \centering
    \includegraphics[width=0.95\columnwidth]{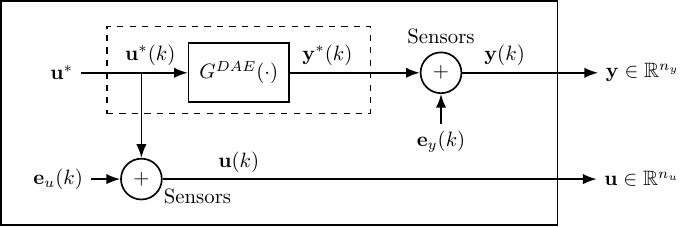}
    \caption{Linear EIV architecture. The true, unobserved inputs and outputs ($\mathbf{u}^*, \mathbf{y}^*$) are corrupted by independent measurement noise ($\mathbf{e}_u, \mathbf{e}_y$) to yield the measured variables ($\mathbf{u}, \mathbf{y}$) consisting of algebraic and differential classes.}
    \label{Figure_1}
\end{figure}

In practical industrial setting, measurements of both input and output variables obtained from operating processes are inevitably contaminated by independent random noise. Identification methods that explicitly account for corrupted measurements on both sides are collectively referred to as errors-in-variables (EIV) approaches~\cite{Kreiberg:2016,Soderstrom:2018}. The EIV formulation for DAE identification presents perhaps the most significant unresolved challenge in this field. As noted by S{\"o}derstr{\"o}m~\cite{Soderstrom:2024a,Soderstrom:2024b}, while PEM and ML can in principle be applied in an errors-in-variables (EIV) setting after appropriate modifications, an explicit and systematic treatment for descriptor systems remains undeveloped. As illustrated in \Cref{Figure_1}, the architecture of an EIV dynamic model for linear descriptor systems aims to identify the true underlying LTI model $G^{DAE}(\cdot)$, that governs the relationship between the noise-free inputs, $\mathbf{u}^*$, and outputs, $\mathbf{y}^*$, relying solely on their noisy measured counterparts, $\mathbf{u}$ and $\mathbf{y}$. The combination of unequal measurement error variances across different sensors (heteroskedasticity) and the necessity to simultaneously untangle complex algebraic and differential constraints creates a statistically formidable identification problem. To the best of our knowledge, no existing method adequately addresses this specific intersection of EIV and MIMO DAE systems.

Among the proposed methods for EIV dynamical systems, the instrumental variable (IV) approach~\cite{Stoica:1995} exploits instruments that are correlated with the true signals but uncorrelated with the measurement noise, effectively neutralizing the bias induced by noisy inputs. A generalized IV framework for MIMO processes was subsequently developed by S{\"o}derstr{\"o}m~\cite{Soderstrom:2011} by integrating IV concepts with bias-compensated least squares (BCLS)~\cite{Sagara:1977,Linden:2012,Khorasani:2017}. While IV methods do not themselves produce estimates of error variances, BCLS directly estimates them to correct the parameter bias that arises from applying ordinary least squares under EIV conditions. Both families of methods, however, require the process orders to be pre-specified or estimated via information criteria such as AIC or BIC, introducing computational burden and the risk of incorrect order selection. For subspace-based methods MOESP~\cite{Verhaegen:1992} and N4SID~\cite{Van:1994}, which were shown to operate under a noise-free input assumption~\cite{Van:1995}, model order selection similarly relies on a combination of a priori singular value thresholding and a posteriori information criteria. Principal component analysis (PCA) and its extensions have emerged as a principled alternative for linear constraint identification in the EIV setting. The Iterative PCA (IPCA) method~\cite{ShankarSir:2008} provides simultaneous model identification and error variance estimation for static processes, exploiting the heteroskedastic noise structure to derive a theoretically motivated order criterion. Its dynamic extension, Dynamic Iterative PCA (DIPCA)~\cite{Maurya:2018}, generalizes this framework to SISO dynamic models, with a subsequent extension to MIMO systems for subspace-based model identification~\cite{Ramnath:2023}. These PCA-based methods represent the most principled EIV framework for dynamic systems, combining statistically consistent parameter estimation with automatic model order determination, without recourse to information criteria such as AIC or BIC.

Beyond the EIV consistency considerations, the identification of each individual DAE equation in its minimal realization form, rather than only recovering a global constraint or transfer function matrix, carries distinct and significant advantages that further motivate the present work. The theory of minimal bases of rational vector spaces \cite{Forney:1975} establishes that polynomial matrix descriptions of multivariable linear systems admit minimal row representations whose invariant structural indices; specifically the observability indices characterize the system's dynamical structure unambiguously~\cite{Wolovich:1974,Morse:1973}. The identification of individual equations in their respective minimal forms directly recovers these invariants, providing a complete and non-redundant characterization of both the differential and algebraic structure of the DAE system. From a statistical standpoint, individual equation identification \cite{Koopmans:1949} affords the flexibility to apply hypothesis tests on specific model coefficients, including tests for coefficient significance, structural adequacy~\cite{Anderson:1949}, and constraint satisfaction, which are not directly accessible from a global constraint matrix. Residual analysis on individually identified equations enables targeted model validation and supports the isolation of misspecified relationships \cite{Bollen:2019}. From a physical perspective, each individual DAE equation typically corresponds to a distinct physical law \cite{Hairer:1996,Ascher:1998}, such as a conservation equation, a constitutive relation, or a kinematic constraint, and this one-to-one correspondence is lost when only the global prediction model is recovered \cite{Unger:1995,Pryce:2001}; the physical content of each law becomes entangled across the rows of the global constraint matrix \cite{Arnold:2017,Iwata:2019}. Collectively, these considerations provide a compelling case for identification methodologies that target individual equation identification, as pursued in the present work.

Despite the advances outlined above, existing identification methodologies for linear descriptor systems exhibit the following fundamental limitations, particularly when confronted with systems of mixed differential-algebraic structure: (i) Existing EIV identification frameworks \cite{Maurya:2018} are formulated for systems with homogeneous ODE structures and cannot accommodate DAE systems containing differential equations of different orders alongside multiple algebraic constraints. No systematic methodology exists for simultaneously determining the number of differential and algebraic equations, the individual equation orders and delay parameters, and the associated model coefficients in their minimal realizable difference-equation form, all from EIV data without prior user specification, (ii) Existing approaches that attempt to extend identification to descriptor systems typically require extensive symbolic manipulation \cite{Brown:1998}, index reduction \cite{Mattsson:1993}, or numerical integration procedures \cite{Brenan:1995}. These requirements render the identification process computationally expensive and numerically fragile for higher-dimensional systems and extended time horizons, precluding scalable and reliable identification of DAE systems with multiple coupled dynamic interactions, and (iii) Current DAE identification methods either ignore the EIV structure entirely, or, when noise in both channels is considered, rely on assumptions of homoskedastic noise. No method has been proposed that simultaneously handles iterative estimation of a diagonal heteroskedastic error covariance matrix and identifies the full mixed differential-algebraic structure of the system without requiring prior specification of model orders or noise levels.

To address each of these limitations, this work presents a systematic methodology for identifying EIV DAE models of linear descriptor systems using a novel DISPCA (Dynamic Iterative-Sequential Principal Component Analysis) framework, with the following primary contributions: (i) A \textit{partial stacking procedure} employing a sequentially increasing lag window that enables the individual identification of each differential relation. This procedure simultaneously determines the number of differential and algebraic equations, the observability indices, and the delay parameters of each differential relation directly from EIV data, without requiring prior specification of model orders from the user, (ii) A systematic decomposition of the DAE identification problem that maintains computational tractability for multi-equation, mixed-order systems. By identifying differential relations sequentially and individually, the methodology avoids symbolic manipulation and index reduction while fully accommodating the complexity of coupled differential-algebraic structures, and (iii) An iterative estimation of the diagonal heteroskedastic measurement error covariance matrix under large sample conditions, which, combined with the sequential identification strategy, provides statistically consistent estimation of all model coefficients for both algebraic and differential components of the DAE system simultaneously. While throughout this work we assume the inputs to be random binary sequences (RBS), a further advantage of the proposed methodology is its flexibility to seamlessly integrate prior knowledge, such as known lag structures, specific differential orders, input variable dynamics, or measurement error variances, if they are available.

The remainder of the paper is structured as follows. \Cref{sec:problem} establishes the system description and formulates the identification problem, detailing the representations of discrete-time ODE MIMO and linear descriptor systems. It also formalizes the concept of individual differential equation orders, defines the measurement error model, and provides the precise problem statement. \Cref{sec:foundation} lays the theoretical groundwork, beginning with IPCA for algebraic constraint identification, specifically detailing the role of scaling in PCA, the iterative estimation of error variances, and the necessary identifiability conditions. This is followed by an examination of DIPCA for identifying SISO dynamic EIV models. In \Cref{sec:dispca}, we present the core methodological contribution of this work, which is a systematic method for identifying EIV DAE model of linear descriptor systems in difference-equation form. Comprehensive simulation studies are presented and discussed in \Cref{sec:results} to demonstrate the effectiveness of the newly introduced methodology. Finally, the paper concludes with few remarks and future directions in \Cref{sec:conclusion}, with supplementary mathematical and technical details provided in the \Cref{app}.

\section{System description and problem formulation} \label{sec:problem}
In this section, we formally describe the class of linear descriptor systems under consideration, and the underlying assumptions. We introduce relevant notations used throughout this work, and formulate the identification problem.

\subsection{Discrete-time ODE MIMO system representation} \label{subsec:2.ode-mimo}
A discrete-time linear dynamical system with multiple inputs and outputs is fundamentally described through a set of linear difference equations that capture the dynamic evolution of the system's output variables. For such a system with $n_u$ inputs and $n_y$ outputs, the input-output representation takes the form:
\begin{equation}
    \label{eq:2.1}
    \mathbf{y}^*(k) + \sum_{\ell=1}^p \mathbf{A}_\ell \mathbf{y}^*(k-\ell) = \sum_{j=D}^s \mathbf{B}_j\mathbf{u}^*(k-j)
\end{equation}
where, $\mathbf{y}^*(k) \in \mathbb{R}^{n_y}$ and $\mathbf{u}^*(k) \in \mathbb{R}^{n_u}$ are the output and input vectors of $k$-th instant, respectively. The superscript $(\cdot)^*$ denotes the true (unobserved) values of the corresponding variables. The matrices $\mathbf{A}_\ell : n_y \times n_y$ and $\mathbf{B}_j : n_y \times n_u$ are the coefficient matrices governing the dynamic relationships. The parameters $p$ and $s$ represent the maximum output and input lags, respectively, whereas $D$ is the delay parameter. \Cref{eq:2.1} can be indexed from $j=0$ by setting $\mathbf{B}_j=\mathbf{0}$ for $0\leq j < D$. Based on this foundational representation, the more general descriptor system formulation is constructed.

\subsection{Discrete-time linear descriptor system} \label{subsec:2.dae}
We consider the discrete-time linear descriptor systems, commonly referred to as differential-algebraic equation (DAE) systems, which represent a broader class of dynamical systems encompassing both differential and algebraic relationships among variables. These systems can accommodate multiple differential relations (dynamic equations) of different orders (discussed in the subsequent sections) and algebraic constraints (static relations) simultaneously, providing a comprehensive framework for modeling physical systems with inherent constraints.

The input-output representation of a linear discrete-time DAE system separates these two components as: (i) the differential (dynamic) equations governing the evolution of dynamic variables:
    \begin{equation}
        \label{eq:2.2}
        \mathbf{y}_d^*(k) + \sum_{\ell=1}^p \mathbf{A}_{d,\ell}^{} \mathbf{y}_d^*(k-\ell) = \sum_{j=0}^s \mathbf{B}_{d,j}^{} \mathbf{u}^*(k-j) 
    \end{equation}
\noindent and (ii) algebraic equations representing static constraints:
    \begin{equation}
        \label{eq:2.3}
        \mathbf{y}_a^*(k) + \mathbf{R}_a \begin{bmatrix} \mathbf{y}_d^*(k) \\ \mathbf{u}^*(k) \end{bmatrix} = \mathbf{0}
    \end{equation}
where, $\mathbf{y}_d^* \in \mathbb{R}^{n_d}$ is the vector of differential output variables, which are governed by dynamic equations with temporal memory, and $\mathbf{y}^*_a \in \mathbb{R}^{n_a}$ is the vector of algebraic output variables that are determined instantaneously by the state of other variables. $\mathbf{u}^* \in \mathbb{R}^{n_u}$ is the vector of input variables, that are assumed to be random stochastic sequences. The matrices $\mathbf{A}_{d,\ell} : n_d \times n_d,$ $\mathbf{B}_{d,j} : n_d \times n_u$ represent the coefficients of differential constraints, while $\mathbf{R}_a : n_a \times (n_d + n_u)$ is the algebraic constraint matrix. Note that $n_y = n_d + n_a$ denotes the total number of measured output variables. 

The structure in \Cref{eq:2.2} is canonical in which each differential equation isolates exactly one differential output variable $y_{d,i}^*(k) \in \mathbf{y}_d^*(k)$ (for $i=1,2,\ldots,n_d$) at the current time instant, while incorporating lagged differential variables and inputs \cite{Koopmans:1949}. This decomposition into differential and algebraic components is a natural property of descriptor systems \cite{Brenan:1995,Kunkel:2024}, and ensures that: (i) the differential equations capture all dynamic evolution of $\mathbf{y}_d(k)$; and (ii) the algebraic variables $\mathbf{y}_a(k)$ are deterministically computed from current-time differential variables and inputs. The nature of this structure avoids overparameterization and enables unique identification of the coefficient matrices.

Since we consider the systems with potentially different orders across differential equations, the matrices $\mathbf{A}_{d,\ell}$ and $\mathbf{B}_{d,j}$ are not necessarily of full row rank. The individual observability indices $\eta_i$ (for $i=1,2,\ldots,n_d$) corresponding to each differential output variable govern the dynamic order of each differential relation, detailed in the following.

\subsection{Observability indices in MIMO systems} \label{subsec:2.order}
For MIMO systems, the concept of equation order requires careful definition. Unlike SISO systems, where the system order is unambiguous, MIMO systems possess multiple output channels, each potentially governed by differential equations of different orders. The orders of each individual differential equation is not uniquely defined in general form, as the system of difference equations can be represented through different linear combinations of the original equations. However, when considering a specific set of differential output variables, the fundamental structural property of such systems is captured by their observability indices, also known as Kronecker indices in the minimal basis sense \cite{Forney:1975}.

\begin{definition} \label{def1}
    \textit{For a linear MIMO system with $n_d$ differential outputs, the observability indices $\eta_1, \eta_2, \ldots, \eta_{n_d}$ are positive integers satisfying the following conditions:}
        \begin{enumerate}
            \item \textit{Each $\eta_i$ represents the minimal order of the differential equation governing the $i$-th output variable, i.e., the smallest possible maximum lag or temporal depth in the dependency structure achievable through linear combinations of the original equations:}
                \begin{align}
                    \eta_i = \min_{\mathbf{T} \in \mathcal{T}}\ \Big( \max\ &\Big\{ l \ \mathbf{|}\ \mathrm{coefficient \ of \ } \mathbf{y}_d(k-l) \mathrm{\ or\ } \nonumber \\ 
                    &\mathbf{u}(k-l) \mathrm{\ in\ } \mathbf{T}_i(\mathcal{E}) \ \text{is non-zero} \Big\} \Big) \label{eq:2.4}
                \end{align}
            \textit{where $\mathcal{E} = \mathcal{E}\left(\mathbf{A}_{d,\ell}, \mathbf{B}_{d,j}, \mathbf{R}_a\right)$ denotes the original equation system, $\mathcal{T}$ is the set of full-rank linear transformations, and $\mathbf{T}_i$ is the $i$-th row of $\mathbf{T}$. This ensures that redundant lags are eliminated while preserving system dynamics.}

            \item \textit{The sum $\sum_{i=1}^{n_d} \eta_i = \eta$, where $\eta$ is the McMillan degree of the system, i.e., the minimal number of dynamical components or states required to realize the system output \cite{Wolovich:1974,Muthairi:2002}.}

            \item \textit{For a specific set of differential output variables, these indices are invariant under similarity transformations and uniquely characterize the system's input-output structure \cite{Popov:1972,Morse:1973}. For descriptor systems with known (pre-specified) input variables, these invariances are specific to the choice of variables belonging to algebraic and differential output classes.}
        \end{enumerate}
\end{definition}

The observability indices correspond precisely to what might be termed ``sub-system orders'' or the ``channel orders'' in coupled differential equation representation. They determine how the overall dynamic complexity (McMillan degree) is distributed across the output channels. For systems in \Cref{eq:2.1,eq:2.2}, the $i$-th differential equation has order $\eta_i$, meaning it involves $\mathbf{y}_d(k-\eta_i)$ as the maximum lagged output term or $\mathbf{u}(k-\eta_i)$ as the maximum lagged input term.

Analogous to \Cref{def1}, the maximal lags $p$ and $s$ in \Cref{eq:2.1,eq:2.2} are defined as:
\begin{align}
    p = \min_{\mathbf{T} \in \mathcal{T}}\ \Big[ \max_{i \in \{ 1,\ldots,n_d\}} \Big( \max\ &\Big\{ l\ \mathbf{|}\ \mathrm{coefficient \ of \ } \mathbf{y}_d(k-l) \nonumber \\
    &\mathrm{\ in\ } \mathbf{T}_i(\mathcal{E}) \ \text{is non-zero} \Big\} \Big) \Big] \label{eq:2.5} \\
    s = \min_{\mathbf{T} \in \mathcal{T}}\ \Big[ \max_{i \in \{ 1,\ldots,n_d\}} \Big( \max\ &\Big\{ l\ \mathbf{|}\ \mathrm{coefficient \ of \ } \mathbf{u}(k-l) \nonumber \\
    &\mathrm{\ in\ } \mathbf{T}_i(\mathcal{E}) \ \text{is non-zero} \Big\} \Big) \Big] \label{eq:2.6}
\end{align}

Therefore, the observability indices relate to these maximum lags as $\eta_i = \max(p_i, s_i)$, where $p_i$ and $s_i$ are the maximal output and input lags, respectively, appearing in the $i$-th differential equation.

\subsection{Measurement error model} \label{subsec:2.me}
In practice, variables are measured with errors rather than their true underlying values. We therefore adopt an EIV framework, shown in \Cref{Figure_1}, that acknowledges uncertainty in the measurements of both input and output variables:
\begin{subequations}
    \label{eq:2.7}
    \begin{align}
        &\mathbf{y}_d(k) = \mathbf{y}_d^*(k) + \mathbf{e}_{\mathbf{y}_d}(k); \quad \mathbf{y}_a(k) = \mathbf{y}_a^*(k) + \mathbf{e}_{\mathbf{y}_a}(k) \\
        &\mathbf{u}(k) = \mathbf{u}^*(k) + \mathbf{e}_{\mathbf{u}}(k)
    \end{align}
\end{subequations}
where $\{\mathbf{y}_d^*, \mathbf{y}_a^*\}, \mathbf{u}^*$ are the noise-free counterparts of the corresponding observed values, denoted by $\{\mathbf{y}_d, \mathbf{y}_a\}, \mathbf{u}$, respectively. $\{\mathbf{e}_{\mathbf{y}_d}, \mathbf{e}_{\mathbf{y}_a}\}, \mathbf{e}_{\mathbf{u}}$ are measurement error vectors reflecting the uncertainty inherent in sensor measurements and data acquisition systems.

We assume that all measurement errors follow a multivariate Gaussian distribution with zero mean and unknown heteroskedastic (unequal) variances, recognizing that different sensors or measurement channels may introduce different levels of measurement uncertainties:
\begin{align}
    \label{eq:2.8}
    \mathbf{e_y}(k) \sim \mathcal{N}\left(\mathbf{0, \Sigma_{e_y}} \right); \quad \mathbf{e_u}(k) \sim \mathcal{N}\left(\mathbf{0, \Sigma_{e_u}} \right) 
\end{align}
where, $\mathbf{e_y}(k) = \begin{bmatrix} \mathbf{e}_{\mathbf{y}_d}(k)^\top & \mathbf{e}_{\mathbf{y}_a}(k)^\top \end{bmatrix}^\top$. The error sequences are assumed to further satisfy the following conditions:
    \begin{itemize}
        \item \textbf{Temporal independence}: $\mathbb{E}\left[\mathbf{e}_{\mathbf{y}}(k_1)\ \mathbf{e}_{\mathbf{y}}^\top(k_2)\right] = \delta_{k_1,k_2}$ $\times\ \mathbf{\Sigma}_{\mathbf{e_y}}$ and $\mathbb{E}\left[\mathbf{e}_{\mathbf{u}}(k_1)\ \mathbf{e}_{\mathbf{u}}^\top(k_2)\right] = \delta_{k_1,k_2} \times \mathbf{\Sigma}_{\mathbf{e_u}}$ $\forall k_1,k_2$, ensuring no temporal correlation in measurement errors across different time instances.

        \item \textbf{Cross-variable independence}: $\mathbb{E}\left[\mathbf{e}_{\mathbf{y}}(k_1)\ \mathbf{e}_{\mathbf{u}}^\top(k_2)\right] = \mathbf{0}$ $\forall k_1,k_2$, ensuring that output and input measurement errors are independent.

        \item \textbf{Orthogonality with true signals}: $\mathbb{E}\left[\mathbf{y}^*(k_1)\ \mathbf{e}_{\mathbf{y}}^\top(k_2)\right]$ $= \mathbf{0},$ $\mathbb{E}\left[\mathbf{u}^*(k_1)\ \mathbf{e}_{\mathbf{u}}^\top(k_2)\right] = \mathbf{0},$ $\mathbb{E}\left[\mathbf{y}^*(k_1)\ \mathbf{e}_{\mathbf{u}}^\top(k_2)\right] = \mathbf{0},$ and $\mathbb{E}\left[\mathbf{u}^*(k_1)\ \mathbf{e}_{\mathbf{y}}^\top(k_2)\right] = \mathbf{0}$ $\forall k_1,k_2$, ensuring that measurement errors are uncorrelated with the true underlying signals for leads and lags.
    \end{itemize}
where, $\delta_{k_1,k_2}$ is the Kronecker delta and $\mathbb{E}[\cdot]$ denotes the mathematical expectation operator. 

\subsection{Problem statement} \label{subsec:2.ps}
Given a sequence of noisy discrete-time measurements of $n_u$ inputs and $n_y$ outputs $\left\{\mathbf{u}(k), \mathbf{y}(k)\right\}$ for $k=1,2,\ldots,N$ from a linear descriptor system described in the \Cref{eq:2.2,eq:2.3} satisfying the above noise assumptions, the discrete-time linear DAE identification problem, addressed in this work, consists of estimating:
    \begin{enumerate}
        \item Measurement error covariance matrices $\mathbf{\Sigma}_{\mathbf{e_y}}$ and $\mathbf{\Sigma}_{\mathbf{e_u}}$,
        \item The number of differential equations $n_d$ and algebraic equations $n_a$,
        \item The observability indices $\eta_i$ for $i=1,2,\ldots,n_d$, and the maximal lags $p$ and $s$,
        \item The model matrices $\mathbf{A}_{d,\ell}\ (\ell=1,2,\ldots,p),$ $\mathbf{B}_{d,j}\ (j=0,1,\ldots,s),$ and $\mathbf{R}_a$.
    \end{enumerate}

All these quantities are to be estimated directly from the measured data without any further prior specification from the user, assuming only the availability of the input and output measurements under the stated error assumptions.

\section{Foundations} \label{sec:foundation}
We review the key identification techniques that form the building blocks of the proposed algorithm. We begin with IPCA \cite{ShankarSir:2008}, a method for identifying linear algebraic constraints among process variables in the EIV framework. As discussed subsequently, this is mathematically equivalent to identifying a MIMO DAE system in which the number of dynamic equations $n_d=0$, i.e., a purely algebraic system. The two central ideas inherited from IPCA, namely, the estimation of error variances from data, and the determination of the number of constraints via a hypothesis test on the singular values of the scaled data matrix, are reused extensively in the proposed framework. We then describe the DIPCA framework \cite{Maurya:2018}, which extends IPCA to dynamic LTI SISO processes through the use of lagged data vectors.

\subsection{IPCA for algebraic constraint identification} \label{subsec:3.IPCA}
Consider $N$ samples of $n$-dimensional noise-free variables $\mathbf{z}_n^*(k) = \begin{bmatrix} z_1(k),\ \ldots,\ z_n(k) \end{bmatrix}^\top$, related by $d$ ($<n$) linear algebraic equations at any sampling instant $k$:
\begin{equation}
    \label{eq:3.1}
    \mathbf{Az}^*_{n}(k) = \mathbf{0}_{d\times 1}, \quad \mathbf{A} \in \mathbb{R}^{d\times n}
\end{equation}
where $\mathbf{A}$ is the constraint matrix. In the EIV setting, only noisy measurements $\mathbf{z}_n(k)$ are available, related to the true values by:
\begin{subequations}
    \label{eq:3.2}
    \begin{align}
        &\mathbf{z}_{n}(k) = \mathbf{z}^*_{n}(k) + \mathbf{e}_{n}(k); \\  
        &\mathrm{where},\ \mathbf{e}_{n}(k) \sim \mathcal{N}(\mathbf{0},\mathbf{\Sigma_e})
    \end{align}
\end{subequations}
with additional noise characteristics:
\begin{subequations}
    \label{eq:3.3}
    \begin{align}
        &\mathbb{E}\left[\mathbf{e}_{n}(k_1)\,\mathbf{e}_{n}^\top(k_2)\right] = \delta_{k_1,k_2}\mathbf{\Sigma_e} \\
        &\mathbb{E}\left[\mathbf{z}^*_{n}(k_1)\,\mathbf{e}_{n}^\top(k_2)\right] = \mathbf{0}, \quad \forall \ k_1,k_2
    \end{align}
\end{subequations}
where $\mathbf{\Sigma_e} = \text{diag}\big(\sigma^2_{e_1},\ldots,\sigma^2_{e_n}\big)$ is  the (generally unknown) diagonal error covariance matrix, assumed independent across variables and time. The true values $\mathbf{z}^*_n(k)$ are assumed to be a deterministic sequence with bounded first and second moments, denoted $\mu_{\mathbf{z}_n^*}$ and $\mathbf{\Sigma}_{\mathbf{z}_n^*}$, respectively.

Defining the true data matrix $\mathbf{Z}_n^* = \begin{bmatrix} \mathbf{z}_n^*(1), \ldots, \mathbf{z}_n^*(N) \end{bmatrix}^\top$ $\in \mathbb{R}^{N\times n}$, and analogously the measured data matrix $\mathbf{Z}_n$, error matrix $\mathbf{E}_n$, \Cref{eq:3.1,eq:3.2} become:
\begin{subequations}
    \label{eq:3.4}
    \begin{align}
        &\mathbf{A}\left(\mathbf{Z}_n^*\right)^\top = \mathbf{0}_{d\times N} \label{eq:3.4a} \\
        &\mathrm{subject \ to \ the \ measurement \ errors} \nonumber \\
        &\mathbf{Z}_n = \mathbf{Z}_n^* + \mathbf{E}_n \label{eq:3.4b}    
    \end{align}
\end{subequations}
implying that the rows of $\mathbf{A}$ span the null space of $(\mathbf{Z}_n^*)^\top$, which has rank $n-d$. Identification of the constraint matrix $\mathbf{A}$ is therefore a null-space estimation problem. This formulation corresponds precisely to the identification of a MIMO DAE system with $n_d=0$, i.e., a system governed entirely by algebraic relations among the variables, with no differential (or equivalently, dynamic difference) equations present. IPCA \cite{ShankarSir:2008} solves this problem in the general heteroskedastic case, where the error variances differ across variables.

\subsubsection{Scaled PCA and the role of the hypothesis test} \label{subsubsec:3.hypo}
The core idea behind IPCA is to scale the data matrix by the inverse square root of the error covariance matrix, i.e., $\mathbf{Z_S}_{n} \triangleq \mathbf{Z}_n\mathbf{\Sigma}_{\mathbf{e}}^{-1/2}$,  so that the noise contribution to the covariance of the scaled data becomes the identity matrix:
\begin{equation}
    \label{eq:3.5}
    \mathbf{\Sigma}_{\mathbf{Z}_{\mathbf{S}n}} = \mathbf{\Sigma}_{\mathbf{Z}_{\mathbf{S}n}^*} + \mathbf{I}_n
\end{equation}
A fundamental consequence of \Cref{eq:3.5} is that the smallest $d$ eigenvalues of $\mathbf{\Sigma}_{\mathbf{Z}_{\mathbf{S}n}}$ are identically unity. This spectral property forms the basis of the \textit{hypothesis test} used to determine the number of constraints $d$, where the null hypothesis that the smallest $d$ singular values of $\mathbf{Z_S}_n/\sqrt{N}$ are equal to one is tested successively for decreasing values of $d$, and $\hat{d}$ is taken as the largest value for which the null hypothesis is not rejected (see \Cref{sec:hypo} for details). The constraint matrix is then recovered via SVD:
\begin{equation}
    \label{eq:3.6}
    \mathrm{svd}\left(\frac{\mathbf{Z_S}_n}{\sqrt{N}}\right) = \mathbf{U}_\mathbf{S}\,\mathbf{S}_\mathbf{S}\,\mathbf{V}_\mathbf{S}^\top \implies \mathbf{\hat{A}} = \left(\mathbf{V}_\mathbf{S}\right)_d^\top \times \mathbf{\hat{\Sigma}_e}^{-1/2}
\end{equation}
where $\left(\mathbf{V}_\mathbf{S}\right)_d$ denotes the $d$ eigenvectors of the sample covariance matrix of $\mathbf{Z_S}_n$ corresponding to the smallest $d$ singular values. The estimated $\mathbf{\hat{A}}$ recovers $\mathbf{A}$  up to a non-singular rotation $\mathbf{T}: d\times d$ (i.e., $\mathbf{\hat{A}} = \mathbf{TA}$), which does not affect the uniqueness of the identified regression model. Specifically, partitioning the $n$ variables into $d$ dependent variables $\mathbf{z}_D(k)$ and $(n-d)$ independent variables $\mathbf{z}_I(k)$:
\begin{subequations}
    \label{eq:3.7}
    \begin{align}
        &\mathbf{z}_{n}(k) =
            \begin{bmatrix}
                \mathbf{z}_{D}(k) \\ \mathbf{z}_{I}(k)
            \end{bmatrix}, \quad
        \mathbf{A} =
            \begin{bmatrix}
                \mathbf{A}_D & \mathbf{A}_I
            \end{bmatrix} \\
        &\implies \mathbf{z}_{D}(k) = \underbrace{-\mathbf{A}_D^{-1}\mathbf{A}_I}_{\mathbf{R}}\,\mathbf{z}_{I}(k), \quad \mathbf{A}_D \in \mathbb{R}^{d\times d}
    \end{align}
\end{subequations}
the regression coefficient matrix $\mathbf{R}$ is invariant to any rotation $\mathbf{T}$, ensuring unique model identification.

When the noise is homoskedastic (i.e., $\mathbf{\Sigma_e} = \sigma_e^2\mathbf{I}_n$ with known $\sigma_e^2$), no scaling is required and PCA directly applies \cite{Wang:2002,Jolliffe:2002,ShankarSir:2015}. IPCA \cite{ShankarSir:2008} generalizes this to the heteroskedastic case, where $\mathbf{\Sigma_e}$ is diagonal with unknown, distinct elements.

\subsubsection{Iterative estimation of error variances} \label{subsubsec:3.ipca-steps}
Since $\mathbf{\Sigma_e}$ is generally unknown, IPCA alternates between two steps until convergence:
\begin{enumerate}[step]
    \item\label{step1} \textbf{Constraint estimation:} Given the current estimate $\mathbf{\hat{\Sigma}}^{(t)}_{\mathbf{e}}$, scale the data and apply SVD to obtain $\mathbf{\hat{A}}^{(t)}$ as per \Cref{eq:3.6}. 
    \item\label{step2} \textbf{Noise covariance update:} Update $\mathbf{\hat{\Sigma}}^{(t+1)}_{\mathbf{e}}$ from the constraint residuals.
\end{enumerate}
Concretely, given $\mathbf{\hat{\Sigma}}^{(t)}_{\mathbf{e}}$:
\begin{subequations}
    \label{eq:3.8}
    \begin{align}
        \mathbf{Z}_{\mathbf{S}_n}^{(t)} &= \mathbf{Z}_n \left(\mathbf{\hat{\Sigma}}_{\mathbf{e}}^{(t)}\right)^{-1/2} \\
        \mathrm{svd}\left(\mathbf{Z}_{\mathbf{S}_n}^{(t)}/\sqrt{N}\right) &= \mathbf{U}_{\mathbf{S}}^{(t)}\,\mathbf{S}_{\mathbf{S}}^{(t)}\,\left(\mathbf{V}_{\mathbf{S}}^{(t)}\right)^\top \\
        \mathbf{\hat{A}}^{(t)} &= \left(\mathbf{V}_{\mathbf{S}}^{(t)}\right)_d^\top \times \left(\mathbf{\hat{\Sigma}}_{\mathbf{e}}^{(t)}\right)^{-1/2}
    \end{align}
\end{subequations}
The constraint residuals at iteration $t$ are: 
\begin{equation}
    \label{eq:3.9}
    \mathbf{r}^{(t)}(k) \triangleq \mathbf{\hat{A}}^{(t)}\mathbf{z}_{n}(k) = \mathbf{\hat{A}}^{(t)}\mathbf{e}_{n}(k)
\end{equation}
where the noise-free term vanishes by \Cref{eq:3.1}. The residual covariance is $\mathbf{\Sigma}^{(t)}_{\mathbf{r}} = \mathbf{\hat{A}}^{(t)} \mathbf{\hat{\Sigma}}_{\mathbf{e}}^{(t)} \left( \mathbf{\hat{A}}^{(t)} \right)^\top$, and the updated $\mathbf{\hat{\Sigma}}_{\mathbf{e}}^{(t+1)}$ is obtained by maximizing the likelihood of the residuals:
\begin{equation}
    \label{eq:3.10}
    \underset{\mathbf{\hat{\Sigma}}_{\mathbf{e}}^{(t)}}{\min}\;\; N\,\log\left|\mathbf{\Sigma}_{\mathbf{r}}^{(t)}\right| + \sum_{k=1}^N \left(\mathbf{r}^{(t)}(k)\right)^\top \left(\mathbf{\Sigma}_{\mathbf{r}}^{(t)}\right)^{-1}\left(\mathbf{r}^{(t)}(k)\right)
\end{equation}
Convergence is monitored via the relative change in the sum of the smallest $d$ eigenvalues across iterations.

\subsubsection{Identifiability condition} \label{subsubsec:3.identi}
A necessary identifiability condition arises because $\mathbf{\Sigma}_{\mathbf{r}}^{(t)}$ (a $d\times d$ symmetric matrix) provides $d(d+1)/2$ independent equations to estimate the $n$ unknown diagonal entries of $\mathbf{\Sigma_e}$. This imposes $d(d+1)/2 \geq n$, which requires a sufficient number of algebraic constraints relative to the number of variables for the error variances to be uniquely identifiable.

\subsection{DIPCA for SISO dynamic EIV model identification} \label{subsec:3.DIPCA}
Dynamic Iterative PCA (DIPCA) \cite{Maurya:2018} extends IPCA to the identification of dynamic LTI SISO process models in the EIV framework, by combining IPCA with the concept of lagged data vectors \cite{Ku:1995}. Consider a deterministic SISO process whose true input $u^*$ and output $y^*$ satisfy the following linear difference equation:
\begin{equation}
    \label{eq:3.11}
    y^*(k) + \sum_{\ell=1}^{\eta_y} a_\ell\,y^*(k-\ell) = \sum_{j=D}^{\eta_u} b_j\,u^*(k-j)
\end{equation}
where $\eta_y$ and $\eta_u$ are the output and input orders, respectively, and $D \leq \eta_u$ is the input-output delay. Define the \textit{equation order} $\eta \triangleq \max(\eta_y,\eta_u)$; when $\eta_y \geq \eta_u$, this coincides with the process order. The EIV objective is to estimate (i) $\eta_y,\eta_u,D$; (ii) the error variances $\sigma^2_{e_y}$ and $\sigma^2_{e_u}$; and (iii) the model coefficients $\{a_\ell\}_{\ell=1}^{\eta_y}$ and $\{b_j\}_{j=D}^{\eta_u}$, from noisy measurements $y(k)$ and $u(k)$.

The key observation underlying DIPCA is that \Cref{eq:3.11} can be interpreted as a static linear constraint among lagged process variables. This motivates the construction of a lagged data matrix $\mathbf{Z}_L$:
\begin{align}
    &\mathbf{z}_{L}(k) = \begin{bmatrix}y(k) & \ldots & y(k-L) & u(k) & \ldots & u(k-L)\end{bmatrix}^\top \nonumber \\
    &\mathbf{Z}_{L} = \begin{bmatrix} \mathbf{z}_{L}(L+1) & \mathbf{z}_{L}(L+2) & \ldots & \mathbf{z}_{L}(N) \end{bmatrix}^\top \label{eq:3.12}
\end{align}
where $L$ is a user-specified lag value chosen to be at least as large as the true equation order $\eta$. When $L>\eta$, excessive stacking introduces additional spurious linear dependencies among the lagged noise-free variables. If the true equation order is $\eta$ and the chosen lag is $L$, the number of linear constraints $d$ among the lagged variables satisfies:
\begin{equation}
    \label{eq:3.13}
    d = L - \eta + 1
\end{equation}
This relationship is central to the order determination procedure, that is after IPCA is applied to $\mathbf{Z}_L$ to obtain the estimate $\hat{d}$ via the singular value hypothesis test, the equation order can be estimated as $\hat{\eta} = L-\hat{d}+1$. 

For the SISO dynamic process, the lagged noise covariance matrix has a structured form with only two distinct unknowns $\sigma^2_{e_y}$ and $\sigma^2_{e_u}$:
\begin{equation}
    \label{eq:3.14}
    \mathbf{\Sigma}_{\mathbf{e},L} =
        \begin{bmatrix}[0.25]
            \sigma_{e_y}^2\,\mathbf{I}_{L+1} & \mathbf{0} \\ \\
            \mathbf{0} & \sigma_{e_u}^2\,\mathbf{I}_{L+1}
        \end{bmatrix}
\end{equation}
This structure reduces the identifiability condition (\Cref{subsubsec:3.identi}) to $d(d+1)/2 \geq 2$, a considerably relaxed requirement compared to the general $n$-variable IPCA case. The optimization problem for updating the noise variances within the IPCA iterations is correspondingly modified:
\begin{equation}
    \label{eq:3.15}
    \underset{\mathbf{\hat{\Sigma}}_{\mathbf{e},L}^{(t)}}{\min}\ (N-L)\,\log\left|\mathbf{\Sigma}_{\mathbf{r}}^{(t)}\right| + \sum_{k=1}^{N-L}\left(\mathbf{r}^{(t)}(k)\right)^\top\left(\mathbf{\Sigma}_{\mathbf{r}}^{(t)}\right)^{-1}\left(\mathbf{r}^{(t)}(k)\right)
\end{equation}
where the factor $N-L$ accounts for the loss of $L$ observations due to stacking, and $\mathbf{\Sigma_r} = \mathbf{\hat{A}}^{(t)} \mathbf{\hat{\Sigma}}^{(t)}_{\mathbf{e},L} \left( \mathbf{\hat{A}}^{(t)} \right)^\top$.

Once $\hat{\eta}$ and the error variances $\hat{\sigma}^2_{e_y}$, $\hat{\sigma}^2_{e_u}$ are obtained the model parameters are identified in a second step. The lagged data matrix is reconfigured with $L=\hat{\eta}$,  ensuring that exactly one linear constraint exists among the lagged noise-free variables, and scaled with $\mathbf{\hat{\Sigma}}_{\mathbf{e},\hat{\eta}}^{-1/2}$:
\begin{subequations}
    \label{eq:3.16}
    \begin{align}
        \mathbf{Z_S}_{\hat{\eta}} &= \mathbf{Z}_{\hat{\eta}}\,\mathbf{\hat{\Sigma}}_{\mathbf{e},\hat{\eta}}^{-1/2} \\
        \mathrm{svd}\left(\mathbf{Z_S}_{\hat{\eta}}/\sqrt{N-\hat{\eta}}\right) &= \mathbf{U_S}_{\hat{\eta}}\,\mathbf{S_S}_{\hat{\eta}}\,\mathbf{V_S}^\top_{\hat{\eta}} \\
        \hat{\boldsymbol{\theta}} &= \left(\mathbf{V_S}_{\hat{\eta}}\right)_1^\top \times \mathbf{\hat{\Sigma}}_{\mathbf{e},\hat{\eta}}^{-1/2}
    \end{align}
\end{subequations}
The eigenvector $\left(\mathbf{V_S}_{\hat{\eta}}\right)_1$ corresponding to the smallest singular value of the covariance matrix of $\mathbf{Z_S}_{\hat{\eta}}$ provides an estimate of the model parameter vector $\boldsymbol{\theta} = \big[ 1\ \ a_1\ \ldots\ a_{\eta_y}\ -b_D\ \ldots\ -b_{\eta_u} \big]^\top$ in the scaled variable space. Scaling back via $\mathbf{\hat{\Sigma}}_{\mathbf{e},\hat{\eta}}^{-1/2}$ recovers $\hat{\boldsymbol{\theta}}$ in the original variable space, and normalization by its first element (coefficient of $y^*(k)$) yields the unique parameter vector. The input-output delay $D$ is implicitly identified in this step, as coefficients corresponding to absent lags are estimated to be statistically indistinguishable from zero.

In the following section, as we shift our discussion from dynamic SISO to MIMO processes, we first discuss the subtle modifications incorporated in the standard DIPCA framework, followed by hierarchical construction of the proposed methodology, that not only discovers all the differential equations, but also all the algebraic constraints in a discrete-time linear descriptor system.

\section{Proposed DISPCA methodology} \label{sec:dispca}
\begin{figure*}[!ht]
    \centering
    \includegraphics[width=\textwidth]{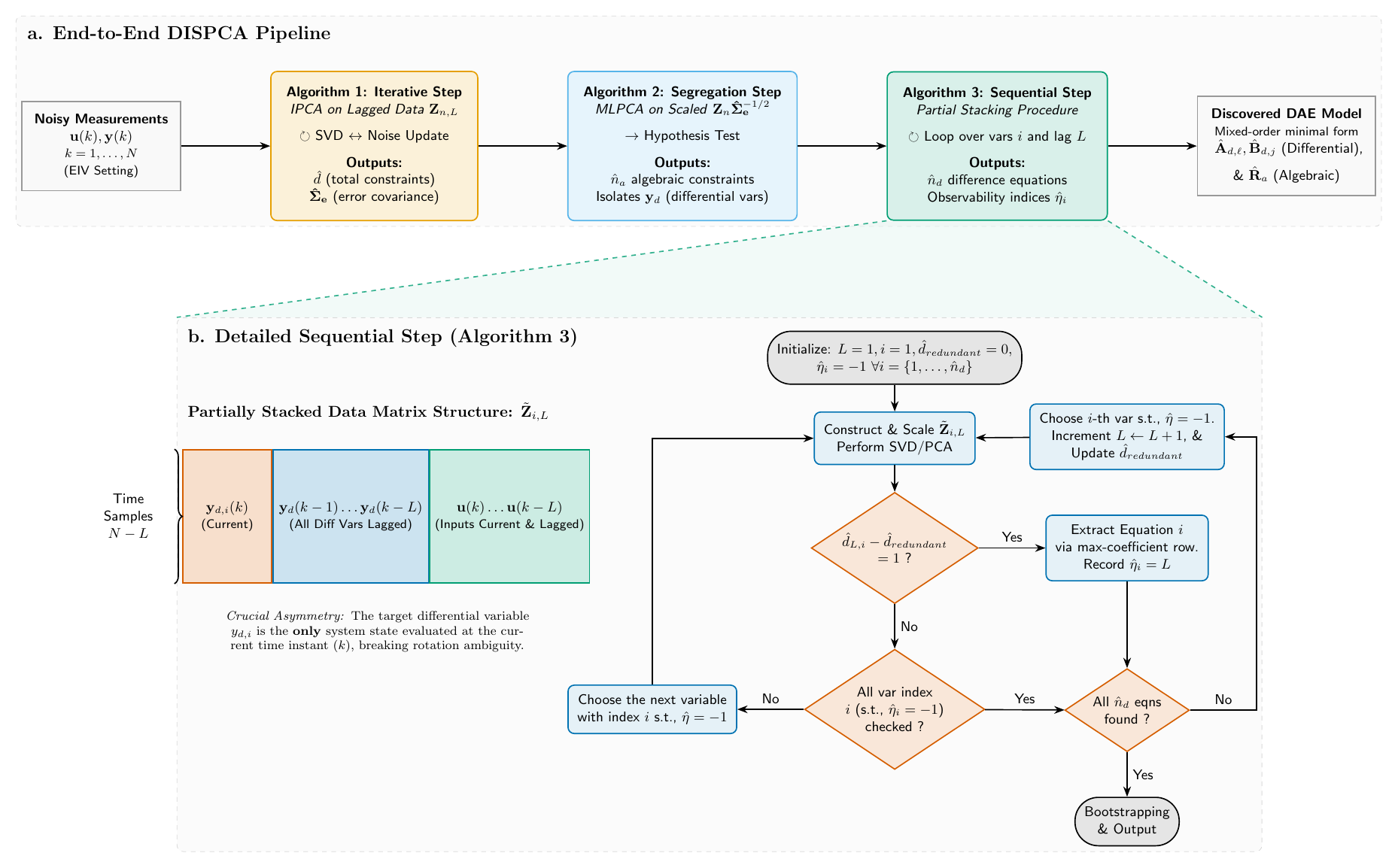}    
    \caption{Schematic overview of the proposed Dynamic Iterative-Sequential Principal Component Analysis (DISPCA) methodology. (a) The macro-level identification pipeline, illustrating the hierarchical progression of noisy measurement data through the Iterative, Segregation, and Sequential steps to extract the mixed-order minimal DAE model. (b) Detailed mechanistic flowchart of the Sequential Step (\Cref{algo:3}). This panel highlights the partial stacking procedure, emphasizing the isolation of the current differential variable to break rotation ambiguity, and the sequential decision loop utilizing the hypothesis test trigger ($\hat{d}_{L,i}-\hat{d}_{redundant}=1$) to uniquely identify individual difference equations.}
    \label{Figure_2}
\end{figure*}

To recapitulate, the complete EIV DAE identification problem involves estimating the following from measurements of $\mathbf{y}^*$ and $\mathbf{u}^*$, denoted as $\mathbf{y}$ and $\mathbf{u}$, respectively:
\begin{enumerate}
    \item Error variances associated with the output and input variable measurements, $\mathbf{\Sigma}_{\mathbf{e}_y}$ and $\mathbf{\Sigma}_{\mathbf{e}_u}$, respectively,

    \item The number of algebraic relations $n_a$, differential relations $n_d$, and the observability index denoted by $\eta_i$, corresponding to $y_{d,i}(k) \in \mathbf{y}_d(k)$,

    \item The model parameters of both the algebraic and differential equations, i.e., $\mathbf{R}_a, \mathbf{A}_{d,\ell},$ and $\mathbf{B}_{d,j}$.
\end{enumerate}
The proposed method, termed as Dynamic Iterative\hyp{}Sequential PCA (DISPCA) which is operated through three successive steps, handles the above identification problem in a hierarchical manner, as illustrated in the schematic pipeline in \Cref{Figure_2}a. First, a modified IPCA is applied to the lagged data matrix $\mathbf{Z}_{n,L}$ constructed as shown in \Cref{eq:4.1} using the combined vector $\mathbf{z}(k) = \begin{bmatrix} \mathbf{y}(k)^\top & \mathbf{u}(k)^\top \end{bmatrix}^\top$. This determines the total number of constraints $\hat{d}$ encoding both the algebraic and differential equations along with their lagged versions, while simultaneously estimating the error covariance matrix $\mathbf{\hat{\Sigma}_e}$ as a byproduct.
\begin{align}
\label{eq:4.1}
    &\mathbf{z}_{L}(k) = 
        \begin{bmatrix}
            \mathbf{z}(k)^\top & \mathbf{z}(k-1)^\top & \mathbf{z}(k-2)^\top & \ldots & \mathbf{z}(k-L)^\top
        \end{bmatrix}^\top \nonumber \\
    &\mathbf{Z}_{n,L} = 
        \begin{bmatrix}
            \mathbf{z}_{L}(L+1) & \mathbf{z}_{L}(L+2) & \ldots & \mathbf{z}_{L}(N)
        \end{bmatrix}^\top  
\end{align}
Subsequently, in the second step, the unlagged data matrix $\mathbf{Z}_n$ is scaled using the estimated error covariance $\mathbf{\hat{\Sigma}_e}$ to identify algebraic constraints $\mathbf{\hat{R}}_a$ and select a possible set of $\hat{n}_a$ algebraic output variables, which isolates the remaining $\hat{n}_d$ differential output variables. Finally, in the third step, we apply a \textit{one-step partial stacking} to each differential variable $y_{d,i} \in \mathbf{y}_d$. The construction of the one-step partially stacked data matrix with respect to the $i$-th differential output variable of $k$-th instant $y_{d,i}(k)$, denoted by $\tilde{\mathbf{Z}}_{i,L}$ follows:
\begin{align}
    \label{eq:4.2}
    &\tilde{\mathbf{z}}_{i,L}(k) = \Big[ y_{d,i}(k) \ \ \mathbf{y}_d(k-1)^\top \ \ldots \ \mathbf{y}_d(k-L)^\top \nonumber \\ &\hspace{7.5em} \mathbf{u}(k)^\top \ \ \mathbf{u}(k-1)^\top \ \ldots \ \mathbf{u}(k-L)^\top \Big]^\top \nonumber \\
    &\tilde{\mathbf{Z}}_{i,L}^{} = \begin{bmatrix}
        \tilde{\mathbf{z}}_{i,L}(L+1) & \tilde{\mathbf{z}}_{i,L}(L+2) & \ldots & \tilde{\mathbf{z}}_{i,L}(N) \end{bmatrix}^\top 
\end{align}
By progressively increasing the lag window $L$ from $1$ and employing bootstrapping \cite{Efron:1994,Robert:2017} until a specific stopping criteria is satisfied, this ensures the discovery of all the differential relations potentially of different orders each corresponding to one of the differential output variables sequentially, while obtaining consistent, unbiased parameter estimates with validated confidence intervals. In the following, we lay out the details of the overall pipeline.

\subsection{Iterative step: Constraint quantification and error covariance estimation} \label{subsec:4.iterative}
We start with estimating the number of linear constraints $\hat{d}$ relating the lagged variables in $\mathbf{Z}_{n,L}$, and in the process, we also obtain the error covariance matrix $\mathbf{\hat{\Sigma}_e}$, that follows the lines of IPCA \cite{ShankarSir:2008,Maurya:2018}. The complete IPCA algorithm is summarized in \Cref{algo:1}, which incorporates the required modifications for dynamic MIMO systems.

\begin{algorithm}[!hbt]
    \caption{\hspace{0.15cm}Iterative estimation of $d$ and $\mathbf{\Sigma_e}$}
    \begin{minipage}{\dimexpr\columnwidth-1\algomargin\relax}
    \vspace{-0.5em}
    \footnotesize
    \KwIn{The data matrix $\mathbf{Z}_n$ and a sufficiently large lag value $L$}
    \KwOut{Number of linear constraints $\hat{d}$, noise covariance matrix $\mathbf{\hat{\Sigma}_e}$}

    \nl Construct $\mathbf{Z}_{n,L}$ using lag $L$ as shown in \Cref{eq:4.1}\;\label{alg1:st1}
    
    \nl Initialize iteration counter $t = 0$ and set $\lambda^{(t=0)}$ to zero. Kick-start the algorithm with the maximum possible guess value for the number of constraints $d_{\mathrm{guess}} = n(L+1)-1$. Let $\mathbf{\hat{\Sigma}}_{\mathbf{e}}^{(t=0)} = \mathbf{I}_n$\;\label{alg1:st2}

    \nl\label{alg1:st3} Construct the estimate of lagged covariance matrix $\mathbf{\hat{\Sigma}}_{\mathbf{e},\mathbf{Z}_{n,L}}^{(t)}$ as\:
        \vspace{-1em}
        \begin{equation}
            \label{eq:4.3}
            \mathbf{\hat{\Sigma}}_{\mathbf{e},\mathbf{Z}_{n,L}}^{(t)} = \mathbf{I}_{L+1}\otimes \mathbf{\hat{\Sigma}}_{\mathbf{e}}^{(t)};
        \end{equation} 
        
    \vspace{-1em}
    \nl Obtain an estimate of constraint matrix $\mathbf{\hat{A}}^{(t+1)}$, given $\mathbf{\hat{\Sigma}}_{\mathbf{e}, \mathbf{Z}_{n,L}}^{(t)}$:
        \vspace{-1em}
        \begin{subequations}
            \label{eq:4.4}
            \begin{align}
                \mathbf{Z}_{\mathbf{S}_{n,L}} &= \mathbf{Z}_{n,L}\mathbf{\hat{\Sigma}}_{\mathbf{e},\mathbf{Z}_{n,L}}^{(t)-1/2} \\
                \mathrm{svd}\left(\mathbf{Z}_{\mathbf{S}_{n,L}}\big/\sqrt{N-L}\ \right) &= \mathbf{U}_{\mathbf{S}_L} \mathbf{S}_{\mathbf{S}_L} \left(\mathbf{V}_{\mathbf{S}_L}\right)^\top \\
                \mathbf{\hat{A}}^{(t+1)} &= \left(\mathbf{V}_{\mathbf{S}_L}\right)_{d_{\mathrm{guess}}}^\top \times \left(\mathbf{\hat{\Sigma}}_{\mathbf{e},\mathbf{Z}_{n,L}}^{(t)}\right)^{-1/2}
            \end{align}
        \end{subequations} 
        where $(\mathbf{V}_{\mathbf{S}_L})_{d_{\mathrm{guess}}}$ denotes the last $d_{\mathrm{guess}}$ columns of $\mathbf{V}_{\mathbf{S}_L}$\;\label{alg1:st4}
    
    \nl Compute $\lambda^{(t+1)}$ as sum of the smallest $d_{\mathrm{guess}}$ singular values contained in $\mathbf{S}_{\mathbf{S}_L}$. Terminate the algorithm if relative change in $\lambda$ falls below specified tolerance threshold and go to step \ref{alg1:st7}; otherwise proceed\;\label{alg1:st5}
    
    \nl Determine the solution for the non-zero elements of $\mathbf{\Sigma_e}$ by minimizing the following function and denote it as $\mathbf{\hat{\Sigma}}_{\mathbf{e}}^{(t+1)}$:
        \begin{align}
            \label{eq:4.5}
                    (N-L) \log\left| \mathbf{\Sigma}_{\mathbf{r},\mathbf{Z}_{n,L}}^{(t)} \right| + \sum_{k=1}^{N-L} \left(\mathbf{r}^{(t)}(k)\right)^\top \left( \mathbf{\Sigma}_{\mathbf{r},\mathbf{Z}_{n,L}}^{(t)} \right)^{-1} \left(\mathbf{r}^{(t)}(k)\right)
        \end{align}
        where $\mathbf{\Sigma}_{\mathbf{r},\mathbf{Z}_{n,L}}^{(t)} = \big(\mathbf{\hat{A}}^{(t+1)}\big) \mathbf{\hat{\Sigma}}_{\mathbf{e},\mathbf{Z}_{n,L}}^{(t)} \big(\mathbf{\hat{A}}^{(i+1)}\big)^\top$\;\label{alg1:st6}
    
    \nl Increment the counter $t$ and repeat steps \ref{alg1:st3} to \ref{alg1:st6} until the convergence\;\label{alg1:st7}
    
    \nl Recover the number of linear constraints $\hat{d}$ from the cardinality of unity eigenvalues using the hypothesis test (refer to \Cref{app}). If $\hat{d} \neq d_{\mathrm{guess}}$, reduce $d_{\mathrm{guess}}$ by $1$ and rerun steps \ref{alg1:st3} to \ref{alg1:st7}\;\label{alg1:st8}
    
    \nl If $\hat{d} = d_{\mathrm{guess}}$, terminate the algorithm and return $\hat{d}$, $\mathbf{\hat{\Sigma}}_{\mathbf{e}}^{(t+1)}$\;\label{alg1:st9}
    \end{minipage}
    \label{algo:1}
\end{algorithm}

It is to note that $\hat{d}$ encodes both the algebraic and differential relations along with their lagged versions. It may also be emphasized that estimation of $\mathbf{\hat{\Sigma}_e}$ from data without any prior knowledge of the model is one of the significant features. Most of the approaches in literature assume that the individual input-output error variances or their ratio is known apriori \cite{Vajk:2005}. The simultaneous estimation of $\hat{d}$ and $\mathbf{\hat{\Sigma}_e}$ leads to a key step, which is the segregation of $n_y$ output variables into possible set of $n_a$ algebraic and $n_d$ differential output variables, as explained in the next subsection.

\subsection{Segregation step: Identification of algebraic constraints} \label{subsec:4.segregation}
The system variables include pre-specified input and output variable classes, where the set of output variables consists of both the algebraic and differential output variable types. Identification of algebraic constraints and thereby the number of algebraic variables follows the discussion from \Cref{subsec:3.IPCA}. Here, we scale the unlagged data matrix (i.e., $\mathbf{Z}_n$) using the estimated error covariance matrix in order to perform the eigenvalue analysis. Since the scaled, unlagged data matrix $\mathbf{Z}_{\mathbf{S}_n}$, which is obtained as:
\begin{equation}
    \label{eq:4.6}
    \mathbf{Z_S}_{n} = \mathbf{Z}_n\mathbf{\hat{\Sigma}}_{\mathbf{e}}^{-1/2}
\end{equation}
does not contain any lagged variables, the null space of $(\mathbf{Z}_n^*)^\top$ is obtained by performing PCA by following \Cref{eq:3.6,eq:3.7}. Let $\hat{n}_a$ be the dimension of the null space in which the algebraic constraint matrix $\mathbf{A}_a$ lies, and therefore the smallest $\hat{n}_a$ eigenvalues should all be unity. This method is also known as maximum likelihood PCA (MLPCA), originally introduced by Wentzell et al.~\cite{Wentzell:1997}. Furthermore, a possible set of $\hat{n}_{a}$ variables can be chosen as algebraic output variables by ensuring the invertibility of the dependent submatrix of $\mathbf{A}_a$. The complete procedure of isolating the algebraic constraints is summarized in \Cref{algo:2}. 

\begin{algorithm}[!hbt]
    \caption{\hspace{0.15cm}Identification of algebraic constraints}
    \begin{minipage}{\dimexpr\columnwidth-1\algomargin\relax}
    \vspace{-0.5em}
    \footnotesize
    \KwIn{Data matrix $\mathbf{Z}_n$ and estimated covariance matrix $\mathbf{\hat{\Sigma}_e}$}
    \KwOut{Algebraic regression matrix $\mathbf{\hat{R}}_a$ and set of $\hat{n}_a$ algebraic outputs}

    \nl Scale the unlagged data matrix $\mathbf{Z}_n$ using $\mathbf{\hat{\Sigma}}_{\mathbf{e}}^{-1/2}$ estimated from \Cref{algo:1}, as shown in \Cref{eq:4.6}. Denote the scaled data matrix as $\mathbf{Z}_{\mathbf{S}_n}$\;\label{alg2:st1}
    
    \nl Apply PCA on $\mathbf{Z}_{\mathbf{S}_n}$ to obtain the algebraic constraint matrix. This includes a hypothesis test to estimate the number of linear constraints.
        \begin{subequations}
            \label{eq:4.7}
            \begin{align}
                \mathrm{svd}\left(\mathbf{Z}_{\mathbf{S}_n}/\sqrt{N}\ \right) &= \mathbf{U_S} \mathbf{S_S} \mathbf{V_S^\top} \\
                \hat{n}_{a} &= \mathrm{hypothesisTest}\left(\mathrm{diag}\left(\mathbf{S_S}\right)^2, N\right) \\
                \hat{\mathbf{A}}_{a} &= \left(\mathbf{V_S}\right)_{\hat{n}_{a}}^\top \times \left(\mathbf{\hat{\Sigma}}_{\mathbf{e}}^{-1/2}\right)
            \end{align}
        \end{subequations}
        where, the function $\mathrm{hypothesisTest}(\cdot)$ (refer to \Cref{app}) takes eigenvalues as input arguments\;\label{alg2:st2}
    
    \nl Partition $\hat{\mathbf{A}}_{a}$ into dependent submatrix $\hat{\mathbf{A}}_{a,D} \in \mathbb{R}^{\hat{n}_a\times \hat{n}_a}$ by choosing $\hat{n}_a$ columns corresponding to $\hat{n}_a$ variables and independent submatrix $\hat{\mathbf{A}}_{a,I} \in \mathbb{R}^{\hat{n}_a\times(n-\hat{n}_a)}$. Ensure the invertibility of $\hat{\mathbf{A}}_{a,D}$ by using a tolerance condition number or determinant as metric values\;\label{alg2:st3}
    
    \nl Identify the unique relations by converting $\mathbf{\hat{A}}_a$ into regression form:
        \begin{equation}
            \label{eq:4.8}
                \hat{\mathbf{A}}_{a} = \begin{bmatrix} \hat{\mathbf{A}}_{a,D} & \hat{\mathbf{A}}_{a,I} \end{bmatrix} \implies \mathbf{\hat{R}}_a = -\big(\hat{\mathbf{A}}_{a,D}^{-1}\big) \hat{\mathbf{A}}_{a,I}
        \end{equation}
        On this constraint matrix $\mathbf{\hat{R}}_a$ in regression form, bootstrapping can further be performed to obtain robust estimates of the parameters\;\label{alg2:st4}
    \end{minipage}
    \label{algo:2}
\end{algorithm}

Thus, the remaining $\hat{n}_d$ ($=n_y-\hat{n}_a$) variables get isolated as possible set of differential output variables, and discovery of corresponding differential relations is discussed below. 

\subsection{Sequential step: Discovery of differential relations} \label{subsec:4.sequential}
This is the final step of our presentation (detailed visually in \Cref{Figure_2}b), where the object of interest is to identify the discrete-time differential relations in difference equation form, each corresponding to one of the $\hat{n}_d$ differential output variables. As per the formulation given in \Cref{eq:2.2}, the differential system requires only the differential output variables and the inputs. Therefore, we remove the set of algebraic variables identified previously (refer to \Cref{subsec:4.segregation}) to construct a reduced system of variables $\tilde{\mathbf{z}}(k) = \begin{bmatrix} \mathbf{y}_d(k)^\top & \mathbf{u}(k)^\top \end{bmatrix}^\top$ of dimension $\tilde{n}\times 1$, where $\tilde{n} = \hat{n}_d+n_u$. It turns out that, we can still use the strength of DIPCA framework before opting for the proposed partial stacking approach that we discuss in the subsequent sections.

\subsubsection{Presence of single differential output variable} \label{subsubsec:4.nd=1}
In presence of a single differential output variable, i.e., $\hat{n}_d=1$, the identification problem essentially corresponds to a multi-input single-output (MISO) dynamical system. Therefore, we can make use of the second step of DIPCA algorithm in order to find the model parameters. With the estimates of total number of constraints $\hat{d}$ relating the variables in $L$-lagged data matrix $\mathbf{Z}_{n,L}$, and the unique algebraic constraints $\hat{n}_a$, the order of that single differential equation can be computed by following \Cref{eq:3.13} as:
\begin{subequations}
    \label{eq:4.9}
    \begin{align}
        &L - \hat{\eta} + 1 = \hat{d} - \hat{n}_a(L+1) \label{eq:4.9a} \\
        \implies &\hat{\eta} = (\hat{n}_a+1)(L+1) - \hat{d} \label{eq:4.9b}
    \end{align}
\end{subequations}
where the RHS of \Cref{eq:4.9a} gives the number of differential constraints relating the variables in $\mathbf{Z}_{n,L}$. Subsequently, the estimation of the model parameter vector follows the discussion in \Cref{subsec:3.DIPCA} (refer to \Cref{eq:3.16}). 

\subsubsection{Presence of multiple differential output variables} \label{subsubsec:4.nd>1}
The specific interest of this discussion is the LTI MIMO dynamical systems where $\hat{n}_d > 1$. Following \Cref{eq:3.13,eq:4.9}, the total number of differential constraints associated with the $L$-lagged data matrix is related to the observability indices of the $\hat{n}_d$ differential relations as:
\begin{subequations}
    \label{eq:4.10}
    \begin{align}
        &\sum_{i=1}^{\hat{n}_d} \left( L - \hat{\eta}_i + 1\right) = \hat{d} - \hat{n}_a(L+1) \\
        \implies &\sum_{i=1}^{\hat{n}_d} \hat{\eta}_i = n_y(L+1)-\hat{d} \label{eq:4.10b}
    \end{align}
\end{subequations}
As a special case, when all the differential relations have identical observability indices, the equation order can be estimated from \Cref{eq:4.10b} as $\hat{\eta}_i = (n_y(L+1)-\hat{d})/\hat{n}_d$. This can further be used to reconstruct a new lagged data matrix with $L=\hat{\eta}_i$ as per the second step of DIPCA to estimate the model parameters. However, in practice, the observability indices corresponding to different differential output variables in a MIMO system could potentially be different, and a standard DIPCA cannot easily be extended in such scenario as evident from \Cref{eq:4.10b}. Since the individual observability indices cannot be extracted, the overall model structure cannot even be constructed, let alone recover the model parameters. 

One essential and fundamental characteristic of the canonical form, given in \Cref{eq:2.2}, of a differential relation with respect to the $i$-th differential output variable (say) $y_{d,i}(k)$ with observability index $\eta_i$ is that the corresponding difference equation includes the lagged versions of all differential output variables, i.e., $\mathbf{y}_d(k^\prime)$ for $k^\prime < k$, and the current ($k$-th) instant of $y_{d,i}$ only. This leads to the idea of \textit{one-step partial stacking}, which is the key contribution of this work. We construct the partially stacked data matrix as defined in \Cref{eq:4.2} using a lag value $L=1$ to kickstart the process. As stated earlier, this data matrix needs to be constructed with respect to each differential output variable one at a time. Therefore, the scaled, one-step partially stacked data matrix with respect to $y_{d,i}(k)$ using lag $L=1$ (depicted in the matrix structure of \Cref{Figure_2}b) is:
\begin{align}
    &\tilde{\mathbf{z}}_{\mathbf{s}_{i,1}}(k) = \begin{bmatrix}
        {y}_{{d,i}_{\mathbf{S}}}(k) & \mathbf{y}_{d_{\mathbf{S}}}(k-1)^\top & \mathbf{u_s}(k)^\top & \mathbf{u_s}(k-1)^\top \end{bmatrix}^\top \nonumber \\
    &\tilde{\mathbf{Z}}_{\mathbf{S}_{i,1}} = \begin{bmatrix}
        \tilde{\mathbf{z}}_{\mathbf{s}_{i,1}}(2) & \tilde{\mathbf{z}}_{\mathbf{s}_{i,1}}(3) & \ldots & \tilde{\mathbf{z}}_{\mathbf{s}_{i,1}}(N) \end{bmatrix}^\top \label{eq:4.11}
\end{align}
where the subscript $(\cdot)_\mathbf{S}$ denotes the scaled versions of these variables obtained by scaling the data using the estimated noise standard deviations. It can be observed from the structure of this data matrix, that it only allows a new differential relation with respect to $y_{{d,i}_\mathbf{S}}(k)$, iff $y_{d,i}^*$ has a first order relation. This can be checked from an eigenvalue analysis after applying SVD on the data matrix and a single unity eigenvalue confirms the existence of the new differential relation corresponding to $y_{d,i}^*(k)$. Furthermore, even if any of the remaining differential output variables have first order relations, the absence of $k$-th instant of these variables in the partially stacked data matrix $\tilde{\mathbf{Z}}_{\mathbf{S}_{i,1}}$ restricts the first order relations from contributing in the number of unity eigenvalues. Therefore the partially stacked data matrix, constructed using $L=1$ with respect to each differential output variable, can be sequentially checked whether there is any further output variable having first order differential relation or not. Once all the $\hat{n}_d$ output variables are sequentially checked and still not all the differential relations are discovered, we increase the lag value $L$ to $2$ and construct the partially stacked data matrix each time with respect to rest of the differential output variables corresponding to which differential relations are yet to be discovered. This can be continued until the existence of all the differential equations is confirmed.

Particularly, at any stage of constructing the partially stacked data matrix using a lag $L$, we need to keep track of the number of redundant relations contributing to the total number of unity eigenvalues. This redundant relations include the differential relations identified from $\ell=1$ up to $\ell=(L-1)$-th stage due to the presence of the lagged variables. Therefore, the number of relations because of the already discovered unique differential equations, i.e., the number of redundant relations can be computed from the following relation, which follows from \Cref{eq:4.10}:
\begin{equation}
    \label{eq:4.12}
    \hat{d}_{redundant} = \sum_{j \neq i;\ y_{d,j} \in\ \mathbf{y}_d} \left((L-1)-\hat{\eta}_j+1\right) 
\end{equation}
where $\hat{\eta}_j$ is the observability index of the differential equation with respect to $y_{d,j}(k)$ discovered in any of the earlier stages $(\ell < L)$, and relation with respect to $y_{d,i}^*(k)$ is yet to be discovered. If $\hat{d}_{L,i}$ denotes the estimated number of unity eigenvalues resulting from the data matrix $\tilde{\mathbf{Z}}_{\mathbf{S}_{i,L}}$, then the existence of the new differential equation corresponding to $y_{d,i}^*(k)$ can be checked by the following decision rule, which serves as the core evaluation loop in our sequential algorithm (see \Cref{Figure_2}b): 
\begin{equation}
    \label{eq:4.13}
    \hat{d}_{L,i} - \hat{d}_{redundant} = 
        \begin{cases}
            1, \ \text{if $\exists$ a new relation} \\[3pt]
            0, \ \text{$\nexists$ new differential relation}
        \end{cases}
\end{equation}

The above equation implies that the estimated observability index corresponding to $y_{d,i}^*(k)$ becomes $\hat{\eta}_i=L$ if a new differential equation is identified. Furthermore, a constraint matrix can be estimated by taking the $\hat{d}_{L,i}$ columns of the right singular matrix $\tilde{\mathbf{V}}_{\mathbf{S}_{i,L}}$, obtained by performing SVD on $\tilde{\mathbf{Z}}_{\mathbf{S}_{i,L}}$, corresponding to the smallest $\hat{d}_{L,i}$ eigenvalues, all of which are approximately equal to unity. An estimate of the $\hat{d}_{L,1} \times \big(1 + \hat{n}_dL + \hat{n}_u(L+1)\big)$ dimensional constraint matrix obtained by rescaling the eigenvectors, i.e., the $\hat{d}_{L,1}$ columns of $\tilde{\mathbf{V}}_{\mathbf{S}_{i,L}}$, is given by:
\begin{align}
    {\mathbf{\hat{A}}}_{\hat{d}_{L,i}} = \left( \tilde{\mathbf{V}}_{\mathbf{S}_{i,L}} \right)_{\hat{d}_{L,i}}^\top \times \left( {\mathbf{\hat{\Sigma}}^{-1/2}_{e,\tilde{\mathbf{Z}}_{i,L}}} \right) \label{eq:4.14} 
\end{align}
\begin{align}
    \text{where}, \mathbf{\hat{\Sigma}}_{e,\tilde{\mathbf{Z}}_{i,L}} = \begin{bmatrix}
        \hat{\sigma}_{e,y_{d,i}}^2 & \mathbf{0} & \mathbf{0} \\
        \mathbf{0} & \mathbf{I}_L \otimes \hat{\mathbf{\Sigma}}_{\mathbf{e},\mathbf{y}_d} & \mathbf{0} \\
        \mathbf{0} & \mathbf{0} & \mathbf{I}_{L+1} \otimes \hat{\mathbf{\Sigma}}_{\mathbf{e},\mathbf{u}}
    \end{bmatrix} \label{eq:4.15}
\end{align}
Here, $\hat{\sigma}_{e,y_{d,i}}^2$ is the estimated variance corresponding to $y_{d,i}$ and the diagonal covariance matrices corresponding to the differential outputs and input variables are denoted as $\hat{\mathbf{\Sigma}}_{\mathbf{e}, \mathbf{y}_d}$, $\hat{\mathbf{\Sigma}}_{\mathbf{e}, \mathbf{u}}$, respectively. The constraint equations relating the noise-free variables corresponding to the stacked data vector $\tilde{\mathbf{z}}_{i,L}(k)$ can be written as \cite{Maurya:2018,Vipul:2020}:
\begin{equation}
    \label{eq:4.16}
    \mathbf{A}_{d_{L,i}} \times \tilde{\mathbf{z}}^*_{i,L}(k) = 
        \begin{bmatrix}
            \mathbf{A}_{\mathbf{y}_d} & -\mathbf{B_u}
        \end{bmatrix}
        \begin{bmatrix}
            \mathbf{y}^*_{i,L}(k) \\
            \mathbf{u}^*_L(k)
        \end{bmatrix} = \mathbf{0}
\end{equation}
where, the constraint matrices $\mathbf{A}_{\mathbf{y}_d}$ and $\mathbf{B_u}$ accounts for the lagged versions of equations in them. 

For instance, consider a system with $2$ differential output variables and $1$ input variable. If the differential equations are of order $1$ and $3$, respectively, then corresponding to the data matrix $\mathbf{z}_{2,3}^*(k)$ which is partially stacked with respect to the second variable having third-order equation, the true structure of $\mathbf{A}_{\mathbf{y}_d}$ and $\mathbf{B_u}$ becomes:
\begin{align}
    &\mathbf{A}_{\mathbf{y}_d} = 
        \begin{bmatrix}[1.35]
            1 & a_1^{(2)} & a_2^{(2)} & a_3^{(2)} & a_4^{(2)} & a_5^{(2)} & a_6^{(2)} \\
            0 & 1 & 0 & a_1^{(1)} & a_2^{(1)} & 0 & 0 \\
            0 & 0 & 0 & 1 & 0 &  a_1^{(1)} & a_2^{(1)}
        \end{bmatrix} \label{eq:4.17} \\
    &\mathbf{B_u} = 
        \begin{bmatrix}[1.35]
            b_0^{(2)} & b_1^{(2)} & b_2^{(2)} & b_3^{(2)} \\ 
            0 & b_0^{(1)} & b_1^{(1)} & 0 \\
            0 & 0 & b_0^{(1)} & b_1^{(1)}
        \end{bmatrix} \label{eq:4.18}
\end{align}
where, the superscript $(\cdot)^{(i)}$ denotes the equation with respect to $i$-th output variable, which the coefficients belong to. However, the estimated dynamic constraint matrix $\mathbf{\hat{A}}_{\hat{d}_{L,i}}$ inherently exhibits rotation ambiguity relative to the true constraint matrix $\mathbf{A}_{d_{L,i}}$, conforming to the relationship:
\begin{equation}
    \label{eq:4.19}
    \begin{bmatrix}
        \mathbf{A}_{\mathbf{y}_d} & -\mathbf{B_u}
    \end{bmatrix} = \mathbf{R}_d\ \times\ \mathbf{\hat{A}}_{\hat{d}_{L,i}}
\end{equation}
where, $\mathbf{R}_d$ represents an unknown rotation matrix. In the special case where $\hat{d}_{L,i} = 1$ and no redundant constraints exist, the constraint vector can be directly normalized by scaling with the inverse of the coefficient corresponding to $y_{d,i}(k)$, yielding the unique difference equation. However, scenarios with $\hat{d}_{L,i} \geq 2$, present apparent complexity due to redundant constraints arising from lagged equation instances. Approaches such as Network Component Analysis (NCA) have been used \cite{Maurya:2022} in order to obtain the elements of $\mathbf{R}_d$ uniquely for SISO systems, where it is noticeable that the determination of first row of $\mathbf{R}_d$ is sufficient to extract the single equation in the presence of its lagged versions. For our mixed-order multi-equation system of interest, we follow a similar approach. In general for such systems, it turns out that, for an equation with observability index $\eta_i$, we require $(\ell_2 - \ell_1 - \eta_i + 1)$ number of constraints to uniquely estimate the equation from $\mathbf{\hat{A}}_{\hat{d}_{L,i}}$ with the known positions of $0$ and $1$ elements in $\mathbf{A}_{\mathbf{y}_d}$ where the variable $y_{d,i}$ has columns of lagged instances from $(k-\ell_1)$ up to $(k-\ell_2)$ in the constraint matrix. This is also the number of rows in which this equation occurs, i.e., the redundancy caused by this equation which can also be confirmed from \Cref{eq:4.17}. This is analogous to the order determination equation in \Cref{eq:3.13} with $\ell_1=0$ and $\ell_2=L$. 

Our fundamental theoretical advancement leverages this result and exploits the intrinsic structure of partially stacked matrices. The construction of $\tilde{\mathbf{Z}}_{\mathbf{S}_{i,L}}$ deliberately isolates $y_{d,i}(k)$ in the first column, ensuring that only the newly discovered dynamic relation involves this variable in the current stage. Since, the variable $y_{d,i}$ is lagged from $(k-0)$-th instant up to $(k-L)$-th instant with a minimal order of $L$, we need $L - 0 - L + 1 = 1$ constraint to uniquely extract this equation from $\mathbf{\hat{A}}_{\hat{d}_{L,i}}$. This is the known ``$1$'' element in the first column and first row of $\mathbf{A}_{\mathbf{y}_d}$. Concurrently, all other rows correspond exclusively to either lagged instances of previously identified equations or linear combinations that exclude $y_{d,i}(k)$. This structural configuration induces a block-triangular form in $\mathbf{R}_d$, where the first row contains a single non-zero entry $[\mathbf{R}_d]_{11}$, while $[\mathbf{R}_d]_{1j} = 0$ for all $j > 1$. Consequently, the target equation can be extracted directly through the row with maximal absolute first coefficient, followed by normalization:
\begin{equation}
    \label{eq:4.20}
        \mathbf{a} = \underset{\forall\ r \ \in\ \{1,\ldots,\hat{d}_{L,i}\}}{\text{argmax}} \left|\mathbf{R}_{r,1}^{}\right|  \implies \mathbf{a}_{\text{norm}} = \mathbf{a} / \mathbf{a}_{1}^{}
\end{equation}
where, $\mathbf{R}_{r,:}$ is the $r$-th row in $\mathbf{\hat{A}}_{\hat{d}_{L,i}}$. This yields the unique difference equation with respect to $y_{d,i}^*(k)$. 

This approach provides significant theoretical and practical advantages over conventional methods. First, it eliminates the requirement for intensive matrix computations by reducing the rotation ambiguity resolution to scalar normalization. Second, sequential lag expansion ensures temporal separation of equations, while the max-coefficient criterion provides inherent numerical stability through preferential selection of well-conditioned constraints. We validate this procedure by retrieving an exact third order equation in presence of multiple first order lagged equations from noise-free data. Consider the following system of equations:
\begin{subequations}
    \label{eq:4.22}
    \begin{align}
        y_1^{*}(k) = +0.70y_1^{*}(k&-1) - 0.02y_3^{*}(k-1) \nonumber \\ &- 0.35u_1^{*}(k-1) - 0.70u_2^{*}(k-1) \label{eq:4.22a} \\
    y_2^{*}(k) = +0.75y_1^{*}(k&-3) \label{eq:4.22b} \\
    y_3^{*}(k) = -0.30y_1^{*}(k&-1) + 0.60y_3^{*}(k-1) \nonumber \\ & + 0.42u_1^{*}(k-1) + 1.10u_2^{*}(k-1) \label{eq:4.22c}
\end{align}
\end{subequations}
The focus is the exact recovery of the equation with respect to $y_2^*(k)$ in presence of lagged instances of the other two first order equations. We generate $4095$ observations using full-band RBS as inputs. Assuming known individual minimal orders, we construct the partially stacked data matrix $\tilde{\mathbf{Z}}^*_{2,3}$ with respect to $y_2^*$ as per \Cref{eq:4.2}. After performing SVD on the true lagged data matrix $\tilde{\mathbf{Z}}^*_{2,3}$, $5$ eigenvalues are identified to be close to zero as reported below:
\begin{align}
    \Lambda = \mathrm{diag}\big(\big[11.5085 \quad &2.8662 \quad \ldots \quad 0.1450 \quad 4.42\times 10^{-4} \nonumber \\ & 9.85 \times 10^{-5} \quad 1.63\times 10^{-31} \nonumber \\ &6.84\times 10^{-32} \quad 4.19\times 10^{-33}\big]\big) \label{eq:4.23}
\end{align}
where, one equation is contributed by $y_2^*(k)$, which accommodates the third order equation and two first order equations are caused by each of $y_1^*$ and $y_3^*$ with respect to their $(k-1)$ and $(k-2)$-th instances. The dynamic constraint matrix $\mathbf{\hat{A}}_5$, obtained by taking the transpose of $5$ columns from $\tilde{\mathbf{V}}_{2,3}$ corresponding to the $5$ zero eigenvalues is:
\begin{align*}
    & \hspace{2em}1 \hspace{4.4em}2 \hspace{1.8em} \ldots \hspace{2.3em}8 \hspace{2.2em} \ldots \hspace{2.5em}18 \nonumber \\
    \mathbf{\hat{A}}_5 = 
        &\begin{bmatrix}
            0.1581 & -0.3714 \!& \ldots & 0.2108 &\! \ldots & -0.6468 \\
            -0.2015 & -0.0592 \!& \ldots & -0.2686 &\! \ldots & -0.2210 \\
            0.0202 & -0.1797 \!& \ldots & -0.0393 &\! \ldots & -0.1017 \\
            0.0557 & -0.6734 \!& \ldots & 0.0244 &\! \ldots & 0.2791 \\
            0.7985 & 0.0274 \!& \ldots & -0.5999 &\! \ldots & -0.0043 \\
        \end{bmatrix}
\end{align*}
Now, as per the above discussion, we choose the fifth row from $\mathbf{\hat{A}}_5$ and after scaling the row using the inverse of first coefficient, we report the coefficients of the third-order equation in \Cref{table:1}. It can be observed that the estimated parameters are in close agreement with the true values. All the coefficients except for $y_2^*(k)$ and $y_1^*(k-3)$ are close to zero (however, they are not exactly zero due to finite-sample errors), which agrees with the true values in \Cref{eq:4.22b}. The coefficients in \Cref{table:1} are reported by considering all the variables to be on the same side of equation, whereas $y_2^*(k)$ is on the opposite side of the equation to maintain a one-to-one correspondence with \Cref{eq:4.22b}.

This concludes the presentation of DISPCA, a self-reliant methodology for identifying mixed-order DAE systems from noisy data. The implicit determination of delay parameters within differential equations is addressed in the following section. For comprehensive implementation details, \Cref{algo:3} specifies the complete sequential estimation procedure.

\begin{table}[htbp]
\centering
\begin{minipage}{\columnwidth}
\centering
\begin{threeparttable}
\caption{Estimated coefficients of the third order system.}
\label{table:1}
\renewcommand{\arraystretch}{1.1}
\begin{tabular}{c@{\hspace{1.25em}}c@{\hspace{1.8em}}c@{\hspace{1.25em}}c}
\toprule
Variable & \makecell{Estimated \\ coefficient} & Variable & \makecell{Estimated \\ coefficient} \\
\toprule
$y_2^*(k)$ & $1.0$ & $y_1^*(k-1)$ & $- 0.0343$ \\
$y_2^*(k-1)$ & $- 1.4102 \times 10^{-16}$ & $y_3^*(k-1)$ & $- 0.0370$ \\
$y_1^*(k-2)$ & $0.0075$ & $y_2^*(k-2)$ & $4.9421 \times 10^{-17}$ \\
$y_3^*(k-2)$ & $0.0131$ & $y_1^*(k-3)$ & $0.7513$ \\
$y_2^*(k-3)$ & $5.9218 \times 10^{-17}$ & $y_3^*(k-3)$ & $0.0049$ \\
$u_1^*(k)$ & $3.5863 \times 10^{-17}$ & $u_2^*(k)$ & $1.4583 \times 10^{-17}$ \\
$u_1^*(k-1)$ & $1.6616 \times 10^{-18}$ & $u_2^*(k-1)$ & $1.6366 \times 10^{-17}$ \\
$u_1^*(k-2)$ & $0.0275$ & $u_2^*(k-2)$ & $0.0167$ \\
$u_1^*(k-3)$ & $0.0054$ & $u_1^*(k-3)$ & $0.0054$ \\
\bottomrule
\end{tabular}
\end{threeparttable}
\end{minipage}
\end{table}

\begin{algorithm}[!hbtp]
    \caption{\hspace{0.15cm}Discovery of differential relations}
    \begin{minipage}{\dimexpr\columnwidth-1\algomargin\relax}
    \vspace{-0.5em}
    \footnotesize
    \KwIn{Scaled unlagged data matrix $\tilde{\mathbf{Z}}_{\mathbf{S}\tilde{n}}$ containing measurements of $\hat{n}_d$ differential output and $n_u$ input variables, covariance matrices $\mathbf{\hat{\Sigma}}_{\mathbf{e},\mathbf{y}_d}, \mathbf{\hat{\Sigma}}_{\mathbf{e},\mathbf{u}}$}
    \KwOut{$\hat{n}_d$ differential equations and individual equation orders $\hat{\eta}$}

    \nl Set the lag $L$ to $1$ and $\hat{d}_{redundant}$ to $0$\;\label{alg3:st1}
    \nl Kick-start the algorithm with a variable index $i=1$\;\label{alg3:st2}
    \nl\label{alg3:st3}\While{$\hat{n}_d$ relations are not yet found} {
        \nl\label{alg3:st4}\If{relation for $y_{d,i}^{}(k)$ is yet to be found} {
            \nl Construct one-step partially stacked data matrix $\tilde{\mathbf{Z}}_{\mathbf{S}_{i,L}}$ with respect to $y_{d,i}(k)$ as shown in \Cref{eq:4.11}\;\label{alg3:st5}
            \nl Perform PCA on $\tilde{\mathbf{Z}}_{\mathbf{S}_{i,L}}$ in order to obtain $\hat{d}_{L,i}$\;\label{alg3:st6}
            \nl\label{alg3:st7}\If{$\hat{d}_{L,i}-\hat{d}_{redundant} = 1$} {
                \nl Construct partially lagged noise covariance matrix $\mathbf{\hat{\Sigma}}_{\mathbf{e}, \tilde{\mathbf{Z}}_{i,L}}$ by following \Cref{eq:4.15}\;\label{alg3:st8}
                \nl Obtain constraint matrix $\mathbf{\hat{A}}_{\hat{d}_{L,i}}$ as per \Cref{eq:4.14}\;\label{alg3:st9}
                \nl Pick the row with highest first max-coefficient and normalize it with the first coefficient (\Cref{eq:4.20})\;\label{alg3:st10}
                \nl Store the difference form of the newly found equation and set its observability index $\hat{\eta}_i$ to $L$\;\label{alg3:st11}    
            }
        }
        \nl\label{alg3:st12}\uIf{$i < \hat{n}_d$} {
            \nl\label{alg3:st13} $i \leftarrow i+1$ \Comment*[r]{Go for the next variable}
        }
        \nl\label{alg3:st14}\ElseIf{$i = \hat{n}_d$} {
            \nl\label{alg3:st15} $i \leftarrow 1$ \Comment*[r]{Reset the variable index}
            \nl\label{alg3:st16} $L \leftarrow L +1$\; 
            \nl Compute $\hat{d}_{redundant}$ as per \Cref{eq:4.12} to consider redundancies from $\ell = 1$ up to $\ell = L$-th stage\;\label{alg3:st17}
        }
    }
    \nl Perform bootstrap to obtain robust parameter estimate and find the delay term in each differential equation\;\label{alg3:st18}
    \nl Use the definitions from \Cref{eq:2.5,eq:2.6} to estimate $\hat{p}$ and $\hat{s}$\;\label{alg3:st19}
    \nl Use the refined parameters to estimate $\mathbf{\hat{A}}_\ell$ and $\mathbf{\hat{B}}_j$ to represent the system of differential equations using \Cref{eq:2.2}\;\label{alg3:st20}
    \end{minipage}
    \label{algo:3}
\end{algorithm}

\section{Case Studies} \label{sec:results} 
We take up two case studies for the demonstration. First study includes a simple RC circuit that results in multiple algebraic and single differential equations. In second study, we discuss a non-interacting three-tank system that produces multiple mixed-order differential and algebraic equations.

\begin{figure}[!htbp]
    \centering
    \includegraphics[width=0.6\columnwidth]{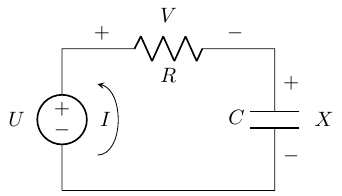}
    \caption{Schematic of the simple RC circuit, highlighting the manipulated input voltage $U$ and the measured variables: voltage $V$ and differential output $X$.}
    \label{Figure_3}
\end{figure}

\subsection{Case study 1: Simple RC circuit} \label{subsec:5.case01}
Consider the RC circuit as shown in \Cref{Figure_3} with a resistor of resistance $R$ and a capacitor with capacitance $C$ driven by a source of voltage $U(t)$. The voltage drops across the resistor and capacitor are denoted as $V(t)$ and $X(t)$, respectively.

\begin{figure*}[t]
    \centering

    \begin{subfigure}[t]{1.85\columnwidth}
        \centering
        \includegraphics[width=\columnwidth,height=0.16\columnwidth]{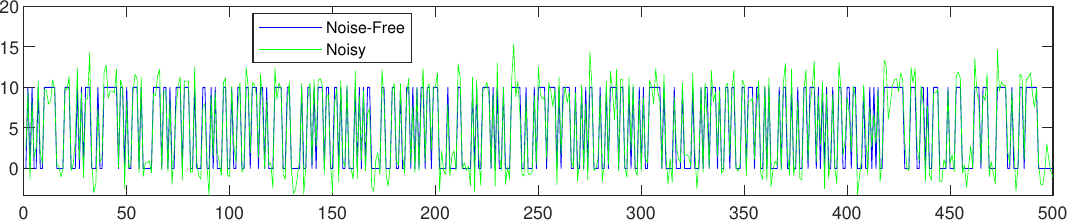}
        \subcaption[Short list entry]{Manipulated input $U(t)$ measurements}
        \label{Figure_4a}
    \end{subfigure}


    \begin{subfigure}[t]{1.85\columnwidth}
        \centering
        \includegraphics[width=\columnwidth,height=0.16\columnwidth]{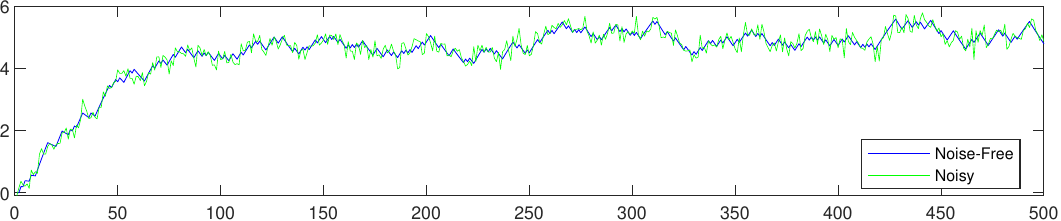}
        \subcaption[Short list entry]{Differential output $X(t)$ measurements}
        \label{Figure_4b}
    \end{subfigure}
    
    \caption{Snapshot of the true and observed input and differential output data as per \Cref{eq:5.1}. The errors in the measurements correspond to a SNR of $10$.}
    \label{Figure_4}
\end{figure*}

If the current across the circuit is assumed to be $I(t)$, the DAE model for this system derived from first principles is:
\begin{align}
    C\frac{dX(t)}{dt} = I(t); \ \ V(t) = RI(t); \ \ U(t) = X(t) + V(t) \label{eq:5.1}
\end{align}
where $X$ is the differential variable, $V$ and $I$ are the algebraic variables. The manipulated input is $U$. The system is simulated using $R = \SI{50}{\ohm}$, $C = \SI{1}{F}$ and excited with a RBS of full bandwidth in order to obtain $4095$ measurements of $U$ which switches between $\SI{0}{V}$ and $\SI{10}{V}$. This is further used to obtain the measurements of $X, V,$ and $I$. The error variances used for simulating the noisy measurements are $\sigma_X^2 = 0.0323,$ $\sigma_V^2 = 2.5448,$ $\sigma_I^2 = 0.0010,$ and $\sigma_U^2 = 2.4986$, which correspond to SNR of $10$. A snapshot of the input $U$ and differential output $X$ data is shown in \Cref{Figure_4}.

\begin{figure}[!htbp]
    \centering
    \includegraphics[width=0.8\columnwidth]{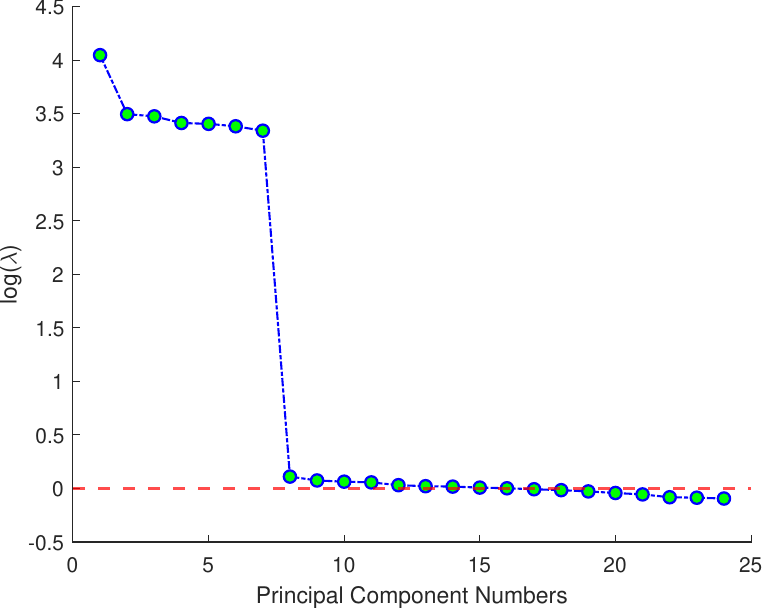}
    \caption{Scree plot associated with the scaled and lagged data matrix $\mathbf{Z_S}_{4,5}$ of the RC circuit. The red dashed line ($y=\log(1) = 0$ line) indicates the threshold for unity eigenvalues used to determine the total linear constraints $\hat{d}$.}
    \label{Figure_5}
\end{figure}

Consider the order of the variables in the data vector be $\mathbf{z}(k) = \begin{bmatrix}X(k) & V(k) & I(k) & U(k)\end{bmatrix}^\top$ and construct the data matrix $\mathbf{Z}_4$ after mean centering the data. A lag $L = 5$ is used to construct the lagged data matrix $\mathbf{Z}_{4,5}$ as per the step \ref{alg1:st1} in \Cref{algo:1}. Following the steps \ref{alg1:st3} to \ref{alg1:st9}, we examine the equality of eigenvalues starting from a maximum possible guess value, $d_{guess} = 23$. While gradually reducing this value, we find that for $d_{guess}=17$, there is an exact match between the guessed number of linear relations and the number of unity eigenvalues, reported in \Cref{Figure_5}. Therefore, it can be concluded that there exist $\hat{d} = 17$ linear relations among the $4(5 + 1) = 24$ lagged variables. The corresponding estimate of diagonal error covariance matrix is obtained as $\mathbf{\hat{\Sigma}}_e = \mathrm{diag}\left(\begin{bmatrix} 0.0331 & 2.4419 & 0.0011 & 2.5179 \end{bmatrix}\right)$, which are fairly close to the true values used. 

\begin{figure}[!htbp]
    \centering
    \includegraphics[width=0.8\columnwidth]{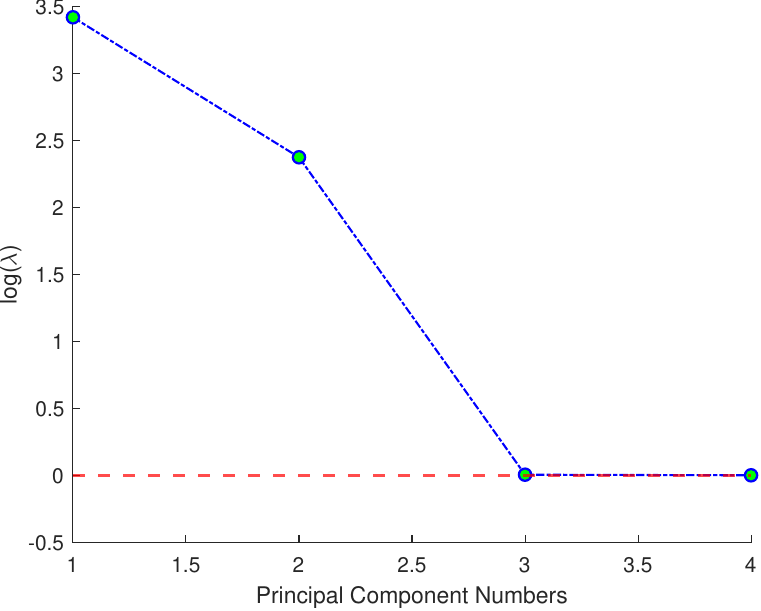}
    \caption{Scree plot associated with the scaled, unlagged data matrix $\mathbf{Z_S}_{4}$ of the RC circuit, used to identify the number of algebraic relations $\hat{n}_a$.}
    \label{Figure_6}
\end{figure}

Further, MLPCA is applied on the unlagged data matrix $\mathbf{Z}_4$ using $\mathbf{\hat{\Sigma}}_{\mathbf{e}}$ as the per steps \ref{alg2:st1} and \ref{alg2:st2} of \Cref{algo:2}. From the reported eigenvalues in \Cref{Figure_6}, it is observed that the number of algebraic relations, $\hat{n}_a = 2$. Therefore, the algebraic constraint matrix $\mathbf{\hat{A}}_a$ is obtained by taking the transpose of the last two columns of the right singular matrix, corresponding to the smallest two eigenvalues. Out of the ${}^3C_2$ combinations of output choices, we choose $z_2$ and $z_3$ as the output variables, corresponding which the dependent submatrix has the maximum determinant value (see \Cref{table:2}). The way, $\mathbf{z}(k)$ is constructed, $z_2$ and $z_3$ correspond to $V$ and $I$, respectively. The estimated algebraic equations are:
\begin{subequations}
    \label{eq:5.2}
    \begin{align}
        &V^*(k) = \underset{\pm(0.0207)}{-0.0098} - \underset{\pm (0.0437)}{0.9761} X^*(k) + \underset{\pm (0.0049)}{0.9898} U^*(k) \label{eq:5.2a} \\
        &I^*(k) = \underset{\pm(0.0047)}{0.0037} - \underset{\pm(0.0009)}{0.0204} X^*(k) + \underset{\pm(0.00008)}{0.0197} U^*(k) \label{eq:5.2b} 
    \end{align}
\end{subequations}
where the estimates of the bias and coefficients are based on averages computed from $200$ runs of bootstrapping and the values in parenthesis are $1.96$ times the standard deviations for the respective parameters, i.e., $95\%$ confidence intervals. 

\begin{table}[!ht]
\centering
\begin{minipage}{\columnwidth}
\centering
\begin{threeparttable}
\caption{Determinant values of the dependent submatrix $\hat{\mathbf{A}}_{a,D}$ \\ corresponding to each output choice for the RC circuit.}
\label{table:2}
\renewcommand{\arraystretch}{1.1}
\begin{tabular}{@{\hspace{3em}}c@{\hspace{7em}}c@{\hspace{3em}}}
\toprule
\makecell{Variable index \\ combinations $i,j$} & $\left|\mathbf{\hat{A}}_{a,D}\right|$ \\
\toprule
$1,$ $2$ & $0.2342$ \\
$1,$ $3$ & $11.2393$ \\
$2,$ $3$ & $11.4701$ \\
\bottomrule
\end{tabular}
\begin{tablenotes}[flushleft]
\footnotesize
\item \hspace*{-0.1cm}Here, $\{i,j\ |\ z_i, z_j \in \mathbf{z}\ \&\ i\neq j \}$, Among all the choices, $z_2,$ $z_3$ are chosen as algebraic output variables for the RC circuit. 
\end{tablenotes}
\end{threeparttable}
\end{minipage}
\end{table}

Since $\hat{n}_d=n_y-\hat{n}_a=1$, we make use of the second step of DIPCA to estimate the parameter vector as discussed in \Cref{subsubsec:4.nd=1} by constructing a data matrix $\tilde{\mathbf{Z}}_{\tilde{n},L'}$ using only the measurements of $X$ and $U$ with a lag value $L'=\hat{\eta}=(2+1)(5+1)-17 = 1$ by following \Cref{eq:4.9b}. This implies that the differential equation with respect to $X^*(k)$ is of first-order. The last column of the right singular matrix, obtained by applying SVD on $\tilde{\mathbf{Z}}_{\mathbf{S}_{\tilde{n},L'}}$ gives the parameters of the difference equation. After scaling it using a properly constructed lagged covariance matrix \Cref{eq:4.15}, the parameters with respect to original variables are obtained as given below:
\begin{align}
    X^*(k) = &\underset{\pm(0.0067)}{0.0013} + \underset{\pm(0.0039)}{0.9795} X^*(k-1)\ - \nonumber \\ 
             &\underset{\pm(0.00092)}{0.00055} U^*(k) + \underset{\pm(0.00052)}{0.0203} U^*(k-1) \label{eq:5.3}
\end{align}

To validate the above result, the true process in the continuous form as in \Cref{eq:5.4a} is discretized using \Cref{eq:5.4b}:
\begin{subequations}
    \label{eq:5.4}
    \begin{align}
        &\dot{\mathbf{x}}(t) = \mathbf{Ax}(t) + \mathbf{Bu}(t) \hspace{1em} \label{eq:5.4a} \\
        &\mathbf{x}(k) = e^{\mathbf{A}\Delta t}\mathbf{x}(k-1)  + \mathbf{A}^{-1}(e^{\mathbf{A}\Delta t} - \mathbf{I})\mathbf{Bu}(k-1) \hspace{1em} \label{eq:5.4b}
    \end{align}
\end{subequations}
where, $\mathbf{x}(k) \triangleq \mathbf{x}(k\Delta t)$ and $\mathbf{x}(t) = X(t)$, $\mathbf{u}(t) = U(t)$. From \Cref{eq:5.1}, after eliminating $I(t)$, the system matrices $\mathbf{A}$ and $\mathbf{B}$ become $-1/RC$ and $1/RC$, respectively. Therefore, this discretization with sampling time $\Delta t = 1$s leads to:
\begin{align}
    &X^*(k) = e^{-1/50} \times X^*(k-1) - (e^{-1/50}-1) \times U^*(k-1) \nonumber \\
    &X^*(k) = 0.9802 \times X^*(k-1)  + 0.0198\times U^*(k-1) \label{eq:5.5}
\end{align}

It is clear that DISPCA has correctly estimated the order, delay and provided accurate estimates of the parameters. 

\begin{figure}[!htbp]
    \centering
    \includegraphics[width=0.9\columnwidth, height=0.75\columnwidth]{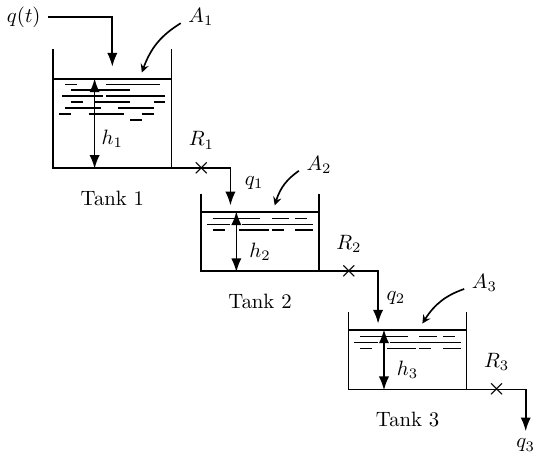}
    \caption{Schematic of the non-interacting three-tank liquid-level system. The inlet flow rate $q(t)$ is the manipulated input, while the tank level $h_3(t)$ and outlet flows ($q_1(t), q_3(t)$) represent the output variables which are measured around the nominal operating point for the current experiment.}
    \label{Figure_7}
\end{figure}

\subsection{Case study 2: Non-interacting three-tank system} \label{subsec:5.case02}
This three-tank system, shown in \Cref{Figure_7}, extends the non-interacting two-tank liquid-level system discussed in \cite{LeBlanc:2009}, which consists of three tanks, namely tank $1$, $2$, and $3$. Here, the outlet flows from tanks $1$ and $2$ discharge directly into tanks $2$ and $3$, respectively. We assume that the liquid to be incompressible, the tanks to have uniform cross-sectional area, and the flow resistances to be linear. Moreover, the liquid flow through valve $R_1$ and $R_2$ depend only on $h_1$ and $h_2$, respectively. The variation in $h_3$ in tank $3$ and $h_2$ in tank $2$ do not affect the transient response occurring in tank $2$ and tank $1$, respectively. Hence, this type of system is known as a non-interacting system. Based on conservation of mass, the deterministic first-principles model of the liquid-level system is as follows:
\begin{subequations}
    \label{eq:5.6}
    \begin{align}
        &q(t) - q_1(t) = A_1\frac{dh_1(t)}{dt}; \\
        &q_1(t) - q_2(t) = A_2\frac{dh_2(t)}{dt}; \\
        &q_2(t) - q_3(t) = A_3\frac{dh_3(t)}{dt} 
    \end{align}
\end{subequations}
The flow-head relationships for the three linear resistances are given by the following expressions:
\begin{align}
    \label{eq:5.7}
    q_1(t) = \frac{h_1(t)}{R_1}; \quad q_2(t) = \frac{h_2(t)}{R_2}; \quad q_3(t) = \frac{h_3(t)}{R_3}
\end{align}

\begin{figure*}[!ht]
    \centering

    \begin{subfigure}[t]{\columnwidth}
        \centering
        \includegraphics[width=0.9\columnwidth,height=0.2\columnwidth]{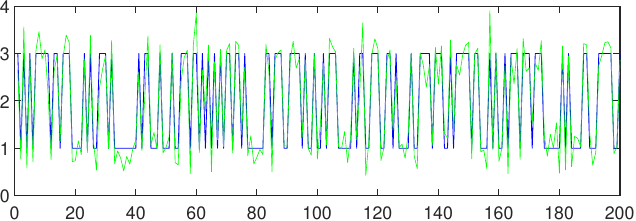}
        \subcaption[Short list entry]{True and observed input data $q(t)$.}
        \label{Figure_8a}
    \end{subfigure}
    \begin{subfigure}[t]{\columnwidth}
        \centering
        \includegraphics[width=0.9\columnwidth,height=0.2\columnwidth]{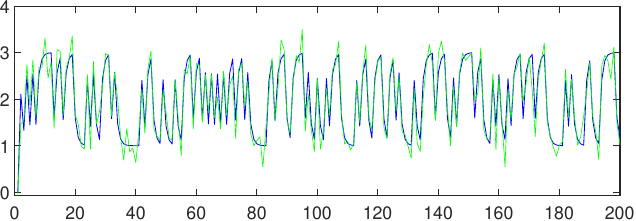}
        \subcaption[Short list entry]{True and observed output data $q_1(t)$.}
        \label{Figure_8b}
    \end{subfigure}

    \vspace{0.8em} 

    \begin{subfigure}[t]{\columnwidth}
        \centering
        \includegraphics[width=0.9\columnwidth,height=0.2\columnwidth]{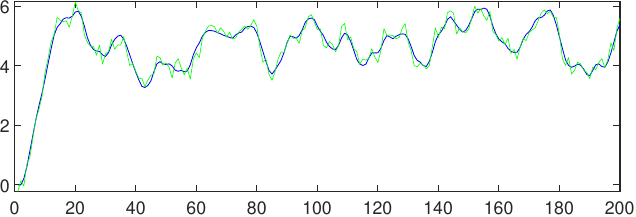}
        \subcaption[Short list entry]{True and observed output data $h_3(t)$.}
        \label{Figure_8c}
    \end{subfigure}
    \begin{subfigure}[t]{\columnwidth}
        \centering
        \includegraphics[width=0.9\columnwidth,height=0.2\columnwidth]{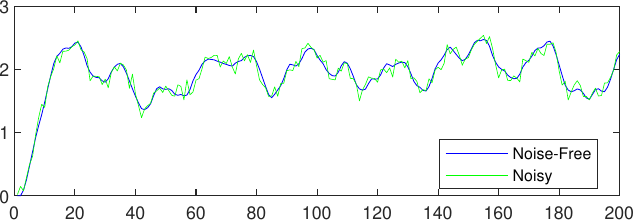}
        \subcaption[Short list entry]{True and observed output data $q_3(t)$.}
        \label{Figure_8d}
    \end{subfigure}
    
    \caption{Snapshot of the true and measured input and output data as per the process defined in \Cref{eq:5.6,eq:5.7}. The measurements correspond to a SNR of $10$.}
    \label{Figure_8}
\end{figure*}

The system is excited with a full-band RBS input $q(t)$ around the nominal operating point, reported in \Cref{table:3}, in order to obtain $4095$ measurements of $q_1(t),\ h_3(t),$ and $q_3(t)$. Gaussian white noise of different variances is added to noise-free data for generating noisy measurements which has a sampling interval of $1$s. The true values of the error variances used for this simulation are provided in \Cref{table:5}, which correspond to a SNR of $10$. A snapshot of the input-output data is shown in \Cref{Figure_8}.

\begin{table}[!htbp]
\centering
\begin{minipage}{\columnwidth}
\centering
\begin{threeparttable}
\caption{Nominal operating conditions and system parameters \\ for the three-tank system presented in \Cref{Figure_7}.}
\label{table:3}
\renewcommand{\arraystretch}{1}
\begin{tabular}{c@{\hspace{2.75em}}c@{\hspace{2.75em}}c@{\hspace{2.75em}}c@{\hspace{2.75em}}c@{\hspace{2.75em}}c}
\toprule
$A_1$ & $A_2$ & $A_3$ & $R_1$ & $R_2$ & $R_3$ \\
\toprule
$1.1$ & $2.4$ & $1.1$ & $0.75$ & $1.5$ & $2.4$ \\
\bottomrule
\end{tabular}
\begin{tablenotes}[flushleft]
\footnotesize
\item \hspace*{-0.1cm}These are the values of system parameters at operating condition of the three-tank system. $A_i\ (\mathrm{m}^2)$ denotes the cross-sectional area of tank $i$, whereas $R_i\ (\mathrm{s}/\mathrm{m}^2)$ is the resistance of the liquid out from tank $i$, for $i=1,2,3$.
\end{tablenotes}
\end{threeparttable}
\end{minipage}
\end{table}

\begin{table}[!htbp]
\centering
\begin{minipage}{\columnwidth}
\centering
\begin{threeparttable}
\caption{Hypothesis testing results for the three-tank case study at \\ a significance level of $\alpha = 0.001$.}
\label{table:4}
\renewcommand{\arraystretch}{1}
\begin{tabular}{c@{\hspace{1.65em}}c@{\hspace{1.65em}}c@{\hspace{1.65em}}c@{\hspace{1.65em}}c}
\toprule
$d_{guess}$ & $df$ & $\tau$ & $\tau_{crit}$ & $H_0$ rejected?  \\
\toprule
$23$ & $275$ & $94423.6717$ & $353.2038$ & True \\
$-$ & $-$ & $-$ & $-$ & $-$ \\
$17$ & $152$ & $23455.8031$ & $211.6200$ & True \\
$16$ & $135$ & $1962.5966$ & $191.5196$ & True \\
$15$ & $119$ & $49.6344$ & $172.4177$ & False \\
\bottomrule
\end{tabular}
\begin{tablenotes}[flushleft]
\footnotesize
\item \hspace*{-0.1cm}These results correspond to $\alpha = 0.001$, and are obtained while performing \Cref{algo:1} in presence of all the variables. Refer to \Cref{app} for more details about the hypothesis test.
\end{tablenotes}
\end{threeparttable}
\end{minipage}
\end{table}

We first apply the \Cref{algo:1} on an excessively stacked data matrix $\mathbf{Z}_{4,L}$ using a lag value $L=5$, where the order of the $n=4$ variables in the data vector is $\mathbf{z}(k) = [q_1(k)\ \ h_3(k)\ \ q_3(k)\ \ q(k)]^\top$. Here, we start with $d_{guess}=23$, which is the maximum possible value and for $d_{guess} = 15$, the test fails to reject the null hypothesis, as the results of the hypothesis test are reported in \Cref{table:4}. Therefore, $\hat{d} = 15$ linear relations exist which constrain the $24$ lagged variables. The estimates of the error variances are can be seen in \Cref{table:5}, which are fairly accurate. 

\begin{table}[!htbp]
\centering
\begin{minipage}{\columnwidth}
\centering
\begin{threeparttable}
\caption{Estimates of noise variances $\sigma^2$ for three-tank case study.}
\label{table:5}
\renewcommand{\arraystretch}{1.15}
\begin{tabular}{c@{\hspace{1.8em}}c@{\hspace{1.8em}}c}
\toprule
Corresponding variable & True value & Estimated value \\
\toprule
$\hat{\sigma}_{q_1}^2$ & $0.0538$ & $0.0542$ \\
$\hat{\sigma}_{h_3}^2$ & $0.0425$ & $0.0415$ \\
$\hat{\sigma}_{q_3}^2$ & $0.0074$ & $0.0075$ \\
$\hat{\sigma}_{q}^2$ & $0.1000$ & $0.0983$ \\
\bottomrule
\end{tabular}
\begin{tablenotes}[flushleft]
\footnotesize
\item \hspace*{-0.1cm}These point estimates correspond to $N=4095$ and $L=5$.
\end{tablenotes}
\end{threeparttable}
\end{minipage}
\end{table}

\begin{figure}[!htbp]
    \centering
    \includegraphics[width=0.85\columnwidth]{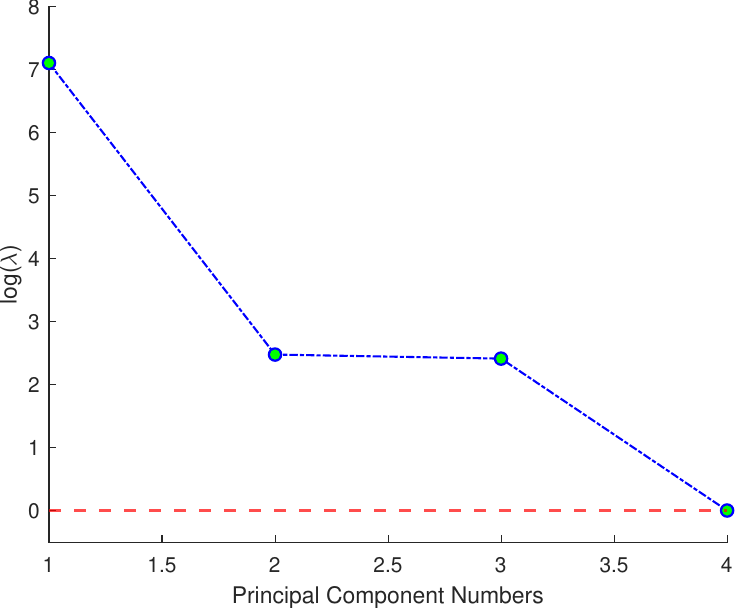}
    \caption{Scree plot associated with $\mathbf{Z_S}_{4}$ corresponding to the three-tank liquid-level system.}
    \label{Figure_9}
\end{figure}

\begin{figure*}[!htbp]
    \centering

    \begin{subfigure}[t]{0.65\columnwidth}
        \centering
        \includegraphics[width=0.95\columnwidth]{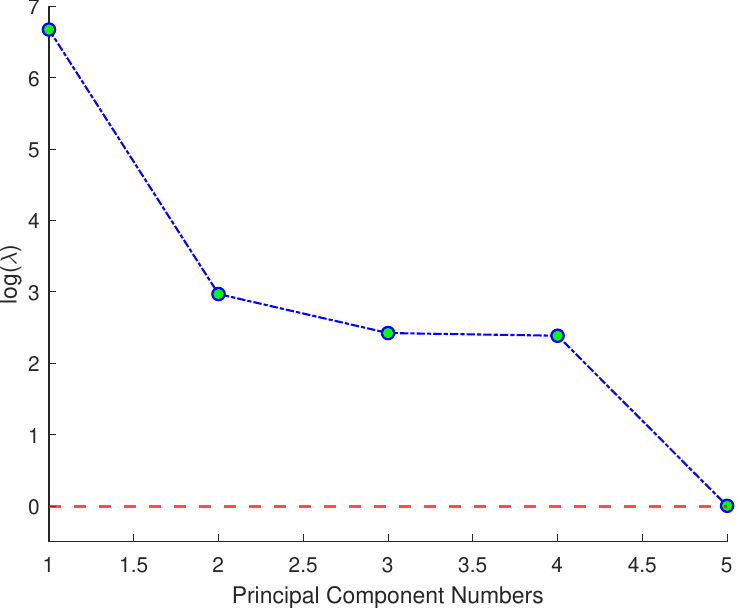}
        \subcaption[Short list entry]{Eigenvalue analysis on $\tilde{\mathbf{Z}}_{\mathbf{S_{1,1}}}$}
        \label{Figure_10a}
    \end{subfigure}
    \begin{subfigure}[t]{0.65\columnwidth}
        \centering
        \includegraphics[width=0.95\columnwidth]{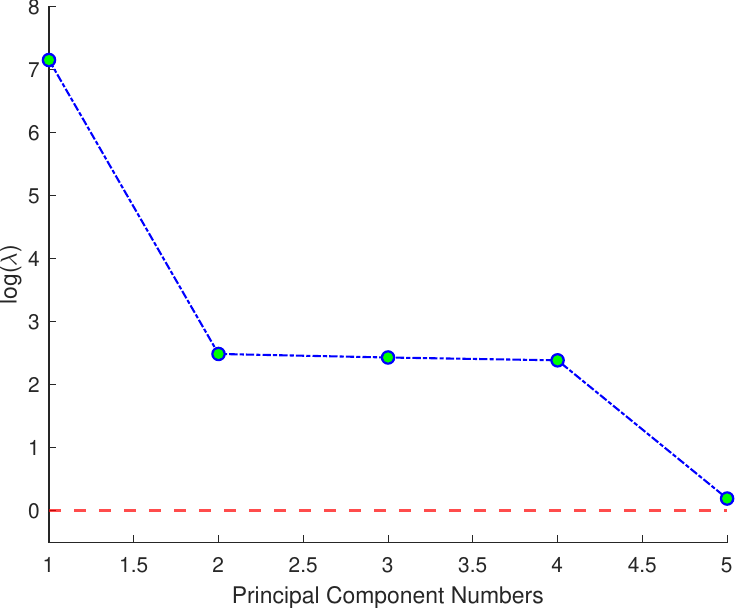}
        \subcaption[Short list entry]{Eigenvalue analysis on $\tilde{\mathbf{Z}}_{\mathbf{S_{2,1}}}$}
        \label{Figure_10b}
    \end{subfigure}
    \begin{subfigure}[t]{0.65\columnwidth}
        \centering
        \includegraphics[width=0.95\columnwidth]{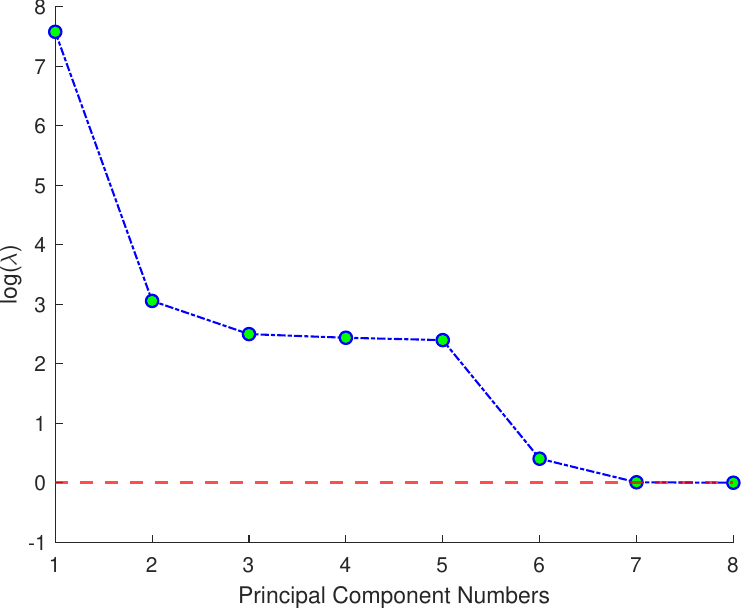}
        \subcaption[Short list entry]{Eigenvalue analysis on $\tilde{\mathbf{Z}}_{\mathbf{S_{2,2}}}$}
        \label{Figure_10c}
    \end{subfigure}
    
    \caption{Scree plots associated with the scaled and partially stacked data matrices in the sequential step (\Cref{algo:3}) for the three-tank system. Panels represent eigenvalue analyses for identifying differential relations corresponding to specific isolated output variables.}
    \label{Figure_10}
\end{figure*}

Subsequently, \Cref{algo:2} results in $\hat{n}_a = 1$ algebraic relation after performing the eigenvalue analysis (refer to \Cref{Figure_9}) on the scaled, unlagged data matrix $\mathbf{Z_S}_4$. Therefore, transpose of the last column of the right singular matrix, corresponding to the smallest eigenvalue, directly gives the algebraic constraint matrix in the scaled domain. $z_3$, which corresponds to $q_3(k)$ is chosen as the algebraic output variable based on the determinant value of the dependent submatrix as reported in \Cref{table:6}. The following is the estimated algebraic equation together with the $1.96$ times of the standard deviation of the respective coefficient estimates, that are shown in parenthesis:
\begin{equation}
    \label{eq:5.8}
    q_3^*(k)  = \underset{\pm(0.0027)}{0.0008}\ q_1^*(k) + \underset{\pm(0.0013)}{0.4168}\ h_3^*(k) + \underset{\pm(0.0019)}{0.0001}\ q^*(k)
\end{equation}
which essentially captures the third flow-head relationship, defined in \Cref{eq:5.7}, accurately. 

\begin{table}[!htbp]
\centering
\begin{minipage}{\columnwidth}
\centering
\begin{threeparttable}
\caption{Determinant values of the dependent submatrix $\hat{\mathbf{A}}_{a,D}$ \\ for algebraic output variable choices in the three-tank \\ system.}
\label{table:6}
\renewcommand{\arraystretch}{1.1}
\begin{tabular}{c@{\hspace{2em}}c@{\hspace{1.5em}}c@{\hspace{1.5em}}c}
\toprule
Variable index $(i)$ & 1 & 2 & 3 \\
\toprule
$\left|\mathbf{\hat{A}}_{a,D}\right|$ & 0.0053 & 3.4403 & 8.2510 \\
\bottomrule
\end{tabular}
\begin{tablenotes}[flushleft]
\footnotesize
\item \hspace*{-0.1cm}The variable indices correspond to the order of variables in $\mathbf{z}(k)$ for the three-tank system. 
\end{tablenotes}
\end{threeparttable}
\end{minipage}
\end{table}

Following the removal of the algebraic output variable $q_3$ from the system, as the last step of DISPCA algorithm, the $\hat{n}_d=2$ differential relations need to identified, which starts with constructing the partially stacked data matrix $\tilde{\mathbf{Z}}_{1,1}$ with respect to $q_1(k)$ using a lag $L = 1$. Eigenvalue analysis on the scaled version this data matrix, as can be seen in \Cref{Figure_10a}, results in $\hat{d}_{1,1} = 1$ unity eigenvalue. Zero redundant relations implies that there exists a first order differential equation with respect to the differential output variable $q_1^*(k)$, which is identified to be as follows:
\begin{align}
    q_1^*(k) =\ &\underset{\pm(0.0044)}{0.3019}\ q_1^*(k-1) - \underset{\pm(0.0026)}{0.0019}\ h_3^*(k-1)\ + \nonumber \\ 
              &\underset{\pm(0.0033)}{0.0007}\ q^*(k) + \underset{\pm(0.0035)}{0.7048}\ q^*(k-1) \label{eq:6.9}
\end{align}
whereas, when the eigenvalue analysis (scree-plot is shown in \Cref{Figure_10b}) is performed on $\tilde{\mathbf{Z}}_{2,1}$, which is constructed with respect to $h_3(k)$, the hypothesis test reports the test statistic and criterion values as $\tau = 9.5146$ and $\tau_{crit} = 3.0902$, i.e., we can reject the null hypothesis (refer to \Cref{subsec:hypo1}). This implies that there is no first order differential equation with respect to $h_3^*(k)$ that relates the variables in the partially lagged data matrix. Therefore, as per the step \ref{alg3:st16} of \Cref{algo:3}, we increase the lag by $1$ and then compute the number of redundant relations. Since, there is only one first-order relation, that has been identified in the previous step, $d_{redundant}$ becomes $(L - 1) = 1$. Subsequently, we find that for $\tilde{\mathbf{Z}}_{2,2}$, constructed with respect to $h_3(k)$, $\hat{d}_{2,2} = 2$, which are also reported in \Cref{Figure_10c}. Therefore, $\hat{d}_{2,2} - \hat{d}_{redundant} = 1$ confirms that there exists a second-order differential equation with respect to $h_3^*(k)$ that relates the variables in $\tilde{\mathbf{Z}}_{\mathbf{S}_{2,1}}$. As per the step \ref{alg3:st10} of \Cref{algo:3}, we pick the row with the maximum first coefficient from the estimated constraint matrix $\mathbf{\hat{A}}_2 \in \mathbb{R}^{2\times 8}$. The estimated difference equation, averaged over $200$ runs of bootstrapping, is as follows:
\begin{align}
    h_3^*&(k) = \underset{\pm(0012)}{0.0701}\ q_1^*(k-1) + \underset{\pm(0.3360)}{1.4309}\ h_3^*(k-1)\ + \nonumber \\
               &\underset{\pm(0.0466)}{0.0894}\ q_1^*(k-2) - \underset{\pm(0.1285)}{0.5198}\ h_3^*(k-2) - \underset{\pm(0.0046)}{0.0084}\ q^*(k)\ + \nonumber \\
               &\underset{\pm(0.0042)}{0.0055}\ q^*(k-1) + \underset{\pm(0.0990)}{0.0237}\ q^*(k-2) \label{eq:6.10}
\end{align}

To validate the above results, the true process described in \Cref{eq:5.6} is discretized under zero order hold (ZOH) assumption. Using only the observable variables used in this case study, the continuous transfer functions in Laplace domain are stated below:
\begin{subequations}
    \label{eq:5.11}
    \begin{align}
        &G_1(s) = \frac{q_1(s)}{q(s)} = \frac{1}{0.825s + 1} \\
        &G_2(s) = \frac{h_3(s)}{q_1(s)} = \frac{2.4}{9.504s^2 + 6.24s + 1}
    \end{align}
\end{subequations}
Discretization of the above transfer functions with sampling time $1$s under ZOH assumption leads to:
\begin{align}
    q_1^*(k) &- 0.2976\ q_1^*(k-1) = 0.7024\ q^*(k-1); \label{eq:5.12} \\
    h_3^*(k) &- 1.4422\ h_3^*(k-1) + 0.5186\ h_3^*(k-2) \nonumber \\
    &= 0.1018\ q_1^*(k-1) + 0.0818\ q_1^*(k-2) \label{eq:5.13}
\end{align}

The estimates derived from DISPCA algorithm are in close agreement to the above discretized transfer functions. This demonstrates the applicability of the DISPCA algorithm to estimate linear DAE models of mixed order systems.

\section{Conclusion and future work} \label{sec:conclusion}
In this work, we proposed the Dynamic Iterative\hyp{}Sequential Principal Component Analysis (DISPCA) methodology for estimating mixed\hyp{}order DAE models of linear descriptor systems. Operating within the EIV framework, this approach successfully identifies systems from data corrupted by uncorrelated, heteroskedastic measurement noise. A primary advantage of DISPCA is its highly automated nature; it systematically executes individual minimal order selection, error covariance estimation, and parameter identification without requiring any further a \textit{priori} specification from the user. Conceptually, DISPCA bridges and generalizes the static-system capabilities of IPCA and the homogeneous\hyp{}order dynamic framework of DIPCA. By utilizing a novel partial stacking procedure, the algorithm robustly identifies underlying models even when the unavailability of specific variable measurements induces mixed\hyp{}order differential relations, which is a practical scenario where traditional methods often fail. Furthermore, the hierarchical structure of DISPCA ensures broad applicability, seamlessly reducing to identify purely algebraic or strictly differential systems depending on the governing physics.

Currently, DISPCA assumes input variables are random sequences and that the measurement error covariance matrix is full rank. Future work will focus on extending the framework to handle input sequences with diverse dynamics, such as auto-correlated inputs, and addressing scenarios involving singular or ill-conditioned error covariance matrices.

\section*{CRediT authorship contribution statement}
\textbf{Deepanjhan Das:} Writing $-$ original draft, Software, Methodology. \textbf{Vishwesh Ramanathan:} Writing $-$ review \& editing, Supervision, Methodology. \textbf{Shankar Narasimhan:} Writing $-$ review \& editing, Supervision, Methodology, Conceptualization.



\appendix
\phantomsection
\refstepcounter{mainappendix} \label{app}
\section*{Appendix}
\addcontentsline{toc}{section}{Appendix}
%
\renewcommand{\thesubsection}{\Alph{subsection}}
\counterwithin{figure}{subsection} 
\counterwithin{table}{subsection}  
\counterwithin{equation}{subsection} 

We present supplementary technical details that underpin the methodology developed in the main text. In \Cref{sec:hypo}, we describe a two-stage sequential testing procedure, first a one-sided test in \Cref{subsec:hypo1} for detecting whether at least one eigenvalue is near unity, then an equality test in \Cref{subsec:hypo2} to estimate how many of the smallest eigenvalues lie close to unity.

\subsection{Hypothesis testing} \label{sec:hypo}
This testing procedure is used in order to obtain the number of linear constraints $d$, corresponding to the number of smallest $d$ eigenvalues of the covariance matrix of the scaled and lagged data matrix $\mathbf{Z_S}_{L}^{}$\footnote{It is used as a general representation of the scaled, lagged data matrix of any of the types (e.g., $\mathbf{Z_S}_{n,L}$, $\mathbf{Z_S}_{n_d(i),L}$) discussed in this work.}, which should all be equal to unity. In order to check for this equality, we perform the hypothesis testing as described in Section 3.7.3 of \cite{Jolliffe:1986}.

\subsubsection{One-sided test} \label{subsec:hypo1}
The first step is to check if there is any unity eigenvalue. In this context, a one-sided test is performed on the smallest eigenvalue $\lambda_P$ to have better flexibility, where $P$ is the total number of eigenvalues. The null and alternate hypothesis of this test are:
\begin{subequations}
    \label{eq:A1}
    \begin{align}
        &H_0: \lambda_P \leq 1 \\
        &H_1: \lambda_P > 1
    \end{align}
\end{subequations}
\noindent A suitable test statistic $(\tau_1)$ is:
\begin{equation}
    \label{eq:A2}
    \frac{\lambda_P-1}{\sqrt{2/(N-1)}},
\end{equation}
which has, approximately, as $\mathcal{N}(0,1)$ distribution under $H_0^{}$, so that $H_0$ will be rejected at significance level $\alpha$ if $\tau_1 > z_\alpha$, which is $100\alpha$ percentile of the standard normal distribution. Throughout this work, we have chosen $\alpha=0.001$.

\subsubsection{Equality test} \label{subsec:hypo2}
If one fails to reject the null hypothesis of the one-sided test, that would imply that $\exists$ at least one unity eigen value. Now, in order to determine how many eigenvalues are equal to unity, we perform the equality test on the $d$ smallest eigenvalues. The null and alternative hypothesis are:
\begin{subequations}
    \label{eq:A3}
    \begin{align}
        &H_{0d}\!: \lambda_{P-d+1} = \lambda_{P-d+2} = \ldots = \lambda_{P} \\
        &H_{1d}\!: \text{at least two of the smallest $d$ $\lambda$'s are different}
    \end{align}
\end{subequations}
The test statistic $\tau_2$ is computed by \Cref{eq:A4} which is as follows:
\begin{align}
    \label{eq:A4}
    \tau_2 = \left(N - \frac{2P+11}{6}\right) \Bigg[ d \ \ln \Bigg( \frac{1}{d} & \sum_{P-d+1}^{P} \lambda_i \Bigg) \nonumber \\ &- \sum_{i=P-d+1}^{P} \ln (\lambda_i) \Bigg]
\end{align}
The test criterion $\tau_{\mathrm{crit}}$ at significance level $\alpha$ is obtained from a $\chi^2_{df,\alpha}$ distribution with degrees of freedom $df$, given by:
\begin{equation}
    \label{eq:A5}
    df = \frac{(d-1)(d+2)}{2}
\end{equation}
The null hypothesis is rejected if $\tau_2 > \tau_{\mathrm{crit}}$. We start from the highest possible value of $d = P-1$ and check till $d = 2$. If the null hypothesis $H_{0d}$ is rejected for all such values of $d$, then $d$ is set to 1, since it fails to reject $H_0$ but rejects $H_{0d} \  \forall d=P-1,\ldots,2$. Subsequently, this implies that $\exists$ only $1$ unity eigenvalue.

\bibliographystyle{elsarticle-num}



\end{document}